\documentclass[english,preprint]{aastex}
\usepackage[T1]{fontenc}
\usepackage[latin9]{inputenc}
\usepackage{color}
\usepackage{rotfloat}
\usepackage{amstext}
\usepackage{amssymb}
\usepackage{graphicx}
\usepackage{textcomp}
\usepackage{gensymb}

\makeatletter

\providecommand{\tabularnewline}{\\}

\makeatother

\usepackage{babel}
\begin{document}

\title{Measurements of Solar Differential Rotation and Meridional Circulation
from Tracking of Photospheric Magnetic Features}

\author{Derek A. Lamb}

\affil{Southwest Research Institute, 1050 Walnut St Suite 300, Boulder,
CO 80302}

\email{derek@boulder.swri.edu}
\begin{abstract}
Long-lived rotational and meridional flows are important ingredients
of the solar cycle. Magnetic field images have typically
been used to measure these flows on the solar surface by cross-correlating
thin longitudinal strips or square patches across sufficiently long
time gaps. Here, I use one month of SDO/HMI line-of-sight magnetic
field observations, combined with the SWAMIS magnetic feature tracking
algorithm to measure the motion of individual features in these magnetograms.
By controlling for perturbations due to short-lived flows and due
to false motions from feature interactions, I effectively isolate
the long-lived flows traced by the magnetic features. This allows
me to produce high-fidelity differential rotation measurements with
well-characterized variances and covariances of the fit parameters.
I find a sidereal rotational profile of $(14.296\pm0.006)+(-1.847\pm0.056)\sin^{2}b+(-2.615\pm0.093)\sin^{4}b$,
with units of $\textrm{ deg d}^{-1}$, and a large covariance $\sigma_{BC}^{2}=-4.87\times10^{-3}(\textrm{ deg d}^{-1})^{2}$.
I also produce medium-fidelity measurements of the much weaker meridional
flow that is broadly consistent with previous results. This measurement
shows a peak flow of $16.7\pm0.6\text{ m s}^{-1}$ at latitude $b=45\degree$ 
but is insufficiently characterized at higher latitudes to ascertain
whether the chosen functional form $2\cos b\sin b$ is appropriate.
This work shows that measuring the motions of individual features
in photospheric magnetograms can produce high precision results in
relatively short time spans, and suggests that high resolution non-longitudinally
averaged photospheric velocity residual measurements could be produced
to compare with coronal results, and to provide other diagnostics
of the solar dynamo.
\end{abstract}


\section{\label{sec:Introduction}Introduction}

Long-lived photospheric flows are important drivers and
diagnostics of the dynamo that is responsible for the solar cycle.
Both flows are now thought to be direct consequences
of the convection that occurs in the outer $\sim$30\% of the Sun
by radius (\citealp[e.g.,][]{Miesch2008_GiantCells}, particularly
\S4). On a global scale the rotational (alternatively longitudinal
or zonal) flow, converts poloidal magnetic field into toroidal magnetic
field. The  meridional (alternatively latitudinal) flow is an essential
component of some tools for studying the solar magnetic cycle that
invoke the Babcock-Leighton mechanism \citep{Babcock1961,Leighton1964,Leighton1969}
such as surface-flux transport simulations \citep[e.g.,][]{Yeates2007_Filament_Surface_Sims,Sheeley2005LRSP}
and flux-transport dynamo models \citep[see][for a recent critical assessment]{Choudhuri2015JApA},
though perhaps not all models require this flow (\citealp{Charbonneau2010LRSP},
\S6.5). Thus understanding the nature of both the rotational and
meridional flows in the solar interior and on the surface is important
for advancing our understanding of the solar dynamo. 

The rotational flow varies with latitude, depth \citep{Thompson1996Sci_GONG_rotation},
and time \citep{1984ApJ...283..385G}; variations in the flow are
strongly linked to the appearance and location of active regions and
the phase of the solar cycle \citep{Howard_&_Labonte1980_Torsional}.
Besides these solar effects, measurements of solar rotation can also
vary for systematic reasons, most prominently the use of different
measurement techniques and methods. \citet{Beck2000SoPh}
provides a review of the measurement techniques and their results
through the end of the 20th century. Spectroscopic observations of
the plasma doppler shift at the limb \citep{1983SoPh_Rotation_Mt._Wilson}
yield different results from observations of surface tracers, and
the rotation curves from different kinds of tracers are as different
as the tracers themselves (a canonical example is \citealp{2000asqu.book.....C},
Table~14.25, and references therein). It is for these reasons that
despite decades (or centuries) of study, what exactly is meant by
the ``solar differential rotation rate'' is subject to considerable
ambiguity. Some of these differences might be attributable to physical
causes such as the different depths at which the tracers are anchored.
Other differences likely arise from poorly understood or uncorrected
systematic effects. Separating physical effects from systematic effects
is important to the interpretation of the measurements: if, for example,
a coronal hole pattern rotates at a different rate than the underlying
magnetic field, this may require an interpretation in which magnetic
reconnection is constantly occurring at the coronal hole boundary;
conversely, if the measured differences arise due to differences in
the definition of the boundary or to difficulties localizing the boundary
in three-dimensional space, then a topologically static coronal hole
boundary may be the best interpretation. Therefore, exploring the
rotational rate measured by different solar features and methods can
still provide useful insight into basic solar processes, nearly 400
years after solar differential rotation was discovered.

The meridional flow also varies with latitude, depth, and time, though
because this flow is 2\textendash 3 orders of magnitude slower than
the rotational flow, its basic structure is still a matter of debate.
While the poleward flow at low and medium latitudes is well-established,
it has not been conclusively settled whether the flow continues all
the way to the pole, or instead a meridional flow counter-cell exists
at high latitudes. The structure of the flow as a function of depth
in the convection zone is even more uncertain, with some measurements
supporting a single cell having a surface poleward flow and a deep
equator-ward return flow, and others supporting more than one cell,
with a surface and deep poleward flows and an equator-ward return
flow at shallower depths in the convection zone \citep{Rightmire2012_HighLatituteMeridional,Zhao2013_Meridional_Double_Cell,Komm2015_Meridional_Solar_Cycle,Rajaguru2015_Meridional_Time-Distance}.

Surface measurements of these flows are accomplished via a variety
of methods: direct measurement of the Doppler shift of a spectral
line, the motion of magnetic field tracers such as sunspots, the small-scale
magnetic pattern, and faculae. The rotation has also been measured
in the corona, using for example coronal holes or coronal bright points.
The values of the rotational and meridional flows produced by a single
method vary over time, which is assumed to be due to solar effects.
However, while there is a lack of agreement between the various methods,
making meaningful comparisons between the different methods is difficult,
because of the solar effects as well as the systematic effects introduced
due to different quantities being measured and different measurement
techniques. One must also consider the possibility that some ``systematic''
effects may in fact be solar, e.g., the difference between the true
rotation rate of sunspots and small magnetic features may well be
solar in origin.

In the past, precise measurements of even the rotational speed typically
required long time periods of measurement, due to a combination of
low data cadence, low spatial resolution, and the need for a statistically
significant number of samples across a wide range of solar latitudes.
A typical technique in photospheric measurements has been to divide
the surface into patches, either roughly square \citep{1990SoPh..130..295H}
or rectangular longitudinal strips \citep{Hathaway2010_MeridionalFlow_SolarCycle},
and perform a cross-correlation of those patches over some time lag
($\sim8$~hrs or $\sim1$~day). Accurately measuring the motion
due to individual elements has been difficult for the same reasons
of cadence, resolution, and statistics. A recent commendable attempt
to acquire a precise measurement of the rotation profile in the solar
corona with only a 2-day dataset, using 906 coronal bright points
as tracers, was made by \citet{Sudar2015}, and was the inspiration
for the present work. However, note that even with wide (5 degree)
latitudinal bins, there was still significant scatter about their
line of best fit (see their Figures~6 \& 7). A much
larger coronal bright point dataset was used by \citet{Sudar2016},
to which I return at the end of the paper.

In this paper I measure the solar photospheric rotational and meridional
flows using measurements of the motion of individual magnetic features.
I use approximately one month of high cadence SDO/HMI observations
of the line-of-sight photospheric magnetic field, and track the positions
of the features using a well-tested feature-tracking algorithm. After
filtering to remove the effect of evolution due to short-lived flows,
I produce measurements of the average rotational and meridional flows
derived from these magnetic features over this one-month time period.
The rotational measurements have a high latitudinal resolution (2\degree)
and are precise, with no significant deviation from the line of best
fit. The values of the fitting parameters approximately agree with
previous work, with well-characterized variances and covariances.
The weaker meridional motion is in rough agreement with previous measurements
of the poleward surface flow, though the measurement is noisier than
the rotational motion when using the same dataset, as expected. The
novel measurement technique presented here opens the future possibility
to measuring non-longitudinally-averaged rotational flows, and high-fidelity
meridional flows with longer data sets.

In \S\ref{sec:Data-Observations} I describe the data and observations,
which includes the methods of measuring the flows. In \S\ref{sec:Results}
I perform fits of analytic functions to both the rotational and meridional
measurements, which includes estimates of the errors in the fit parameters.
In \S\ref{sec:Conclusions} I present my conclusions, including the
possibility of future measurements using this technique.

\section{\label{sec:Data-Observations}Data \& Observations}

I used a series of line-of-sight magnetograms from the Helioseismic
and Magnetic Imager \citep{Scherrer2012HMI} aboard the Solar Dynamics
Observatory \citep{PesnellSDO2012}. The magnetograms were retrieved
from the near-real-time 45~s magnetogram data series at a cadence
of 12~min. The dataset comprises 3000 images beginning at 2011-02-01T00:00:45
TAI and ending at 2011-02-26T08:24:45 TAI. There are 23 instances
of missing frames in the dataset: 17 are gaps of a single frame, and
the remainder are gaps of 2, 2, 4, 5, 5, and 8~frames. I address
the issue of data gaps later in this section.

\begin{figure}
\begin{centering}
\includegraphics[width=0.5\textwidth]{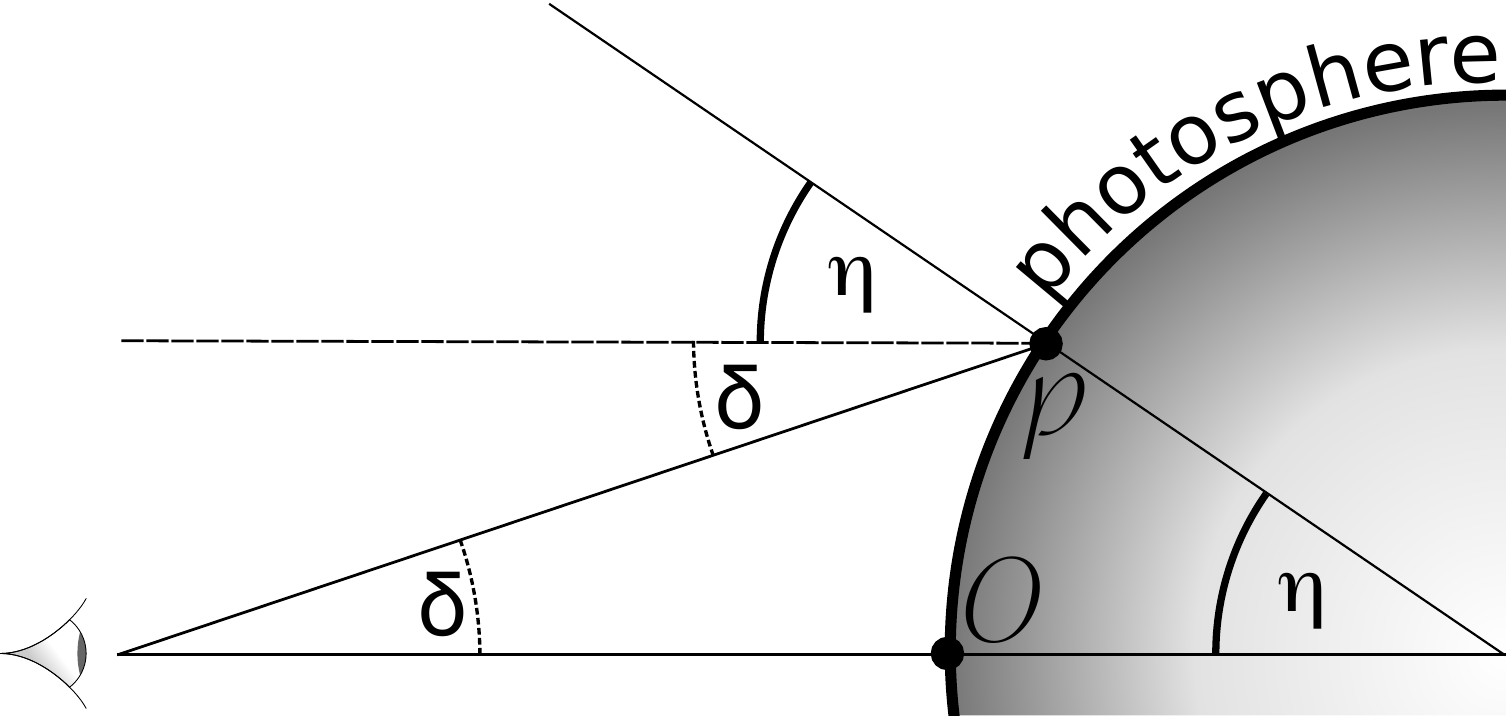}
\par\end{centering}
\caption{\label{fig:eta-geometry}Illustration of the line-of-sight correction
geometry. The angle $\eta$ subtends the photospheric arc connecting
the sub-observer point $\mathcal{O}$ and a given location $\mathcal{P}$
as viewed from the center of the Sun. The angle $\delta$ subtends
these same two points as viewed from the observer. $\alpha=\eta+\delta$
is the angle between local vertical and the line connecting the observer
and $P$. The drawing is not to scale: for an observer at 1~AU $\delta\lessapprox0.25\degree$
and $0\leq\eta\lessapprox90\degree$. The corrections to produce
the \emph{field} and \emph{flux }images involve dividing the input
magnetogram by $\cos(\alpha)$ and $\cos^{2}(\alpha)$, respectively.}

\end{figure}

Preprocessing includes rotation of each image by $180\degree$
so that solar north is ``up'', and for reasons of computational
efficiency reducing the spatial size of the original data by a factor
of 4. Size reduction is accomplished using an optimized filter with
a Hanning ($\cos^{2}$) rolloff \citep{DeForest_resampling}. The
reduced images are $1024\times1024$ pixels, with an angular scale
of 2.0'' per pixel, which is sufficient for these purposes. The input
magnetograms were corrected for magnetic field vector line-of-sight
effects, assuming that the magnetic field is vertical at the photosphere,
a small variation on the technique first used by \citet{2001ApJ...555..448H}.
This is an acceptable assumption away from active regions where I
perform most of the analysis in this paper (but see \citealp{Leka2013_Pre-emerging1}
for an alternative correction that assumes the fields are potential).
For simplicity, I do not use vector magnetogram information in the
active regions\textemdash the contribution from active regions to
my measurements turns out to be nearly negligible. The line-of-sight
correction takes the form of dividing the line-of-sight flux density
in each pixel by $\cos(\alpha)$, $\alpha=\eta+\delta,$ where $\eta$
is the Sun-centered angle subtended by the great circle segment connecting
each pixel to the disk image center, and $\delta$ is the (small)
angle between a line connecting a given pixel to the observer and
a line from that pixel that is parallel to the line connecting Sun
center to the observer (Figure~\ref{fig:eta-geometry}).
Assuming the magnetic field vector is vertical in the region of the
photosphere where the magnetograms are observed, division by $\cos(\alpha)$
fully corrects for the discrepancy between local vertical and the
line-of-sight. The image produced by this correction is called the
\emph{field} image. The field image is again divided by $\cos(\alpha)$
to correct for foreshortening: the increase between disk center and
limb in the solar surface area subtended by a pixel. The image produced
by this correction is called the \emph{flux} image. The \emph{field}
and \emph{flux} images are cropped so that all pixels for which $\alpha\ge70\degree$
are ignored in subsequent processing.

The prepared data were then processed by the Southwest
Automatic Magnetic Identification Suite (SWAMIS) feature tracking
algorithm \citep{2007ApJ...666..576D}, and I refer the reader to
that work for a detailed description of the methodology and terminology.
SWAMIS is five-step algorithm that identifies magnetic features on
the solar surface and associates them across multiple images. Magnetic
features are regions of the solar surface, grouped into collections
of adjacent pixels that all have a pixel-averaged field strength above
a certain low threshold, and that have at least one pixel above a
certain high threshold at some point in time. A feature must also
satisfy certain nominal minimum size and lifetime requirements in
order to make the detections robust against different sources of noise.

In the discrimination step, I used a high detection threshold of 53~G
and a low detection threshold of 15~G on the \emph{field} images.
In the identification step, I used a hybrid combination of the ``downhill''
and ``clumping'' methods, with the clumping method applied to pixels
above a 33~G threshold, and the downhill method applied to pixels
below that threshold. I measured the noise in the scaled-down magnetograms
\citep[as in][]{1999ApJ...511..932H} to be approximately 5.2~G,
so these three thresholds correspond to approximately 10, 3, and 6.5~$\sigma$
respectively.

Association of features between sequential \emph{flux }images was
performed with the dual-maximum flux-weighted overlap method described
by \citep{2007ApJ...666..576D}. The association algorithm makes the
identification number of each feature unique across the dataset, which
enables my image-to-image velocity calculations. It is important to
note that the association algorithm employed here makes no attempt
to account for rotational motion when associating features between
images, so understanding the performance of this step across data
gaps is important. I made a visual comparison between identified features
before and after various-sized gaps, and found that the association
step performs acceptably well on my scaled-down data over data gaps
$\le$ 8~images (96~min). Over larger data gaps the rotational motion
moves features too far for direct association to be accurate, and
the \emph{mistaken identity problem} manifests\textendash a feature
will be incorrectly associated across the gap if an unrelated feature
of the same polarity happens to be present at the same location after
the data gap. After the association step, the tabulation step accumulates
data for each feature in each image into a \emph{feature table} that
contains the flux, area, flux-weighted centroid location, and the
identification number of features that are considered to be neighbors
of this feature, among other quantities. The neighbors of a feature
are those features found within a two-pixel boundary around each feature.
If a feature has neighbors in a given image, I set a \emph{closeness}
parameter to 1 for that feature in that image. After the \emph{feature
table} has been made, a \emph{summary} \emph{table} is made that has
summary information for each feature across all images: birth and
death images numbers, lifetime, and the \emph{average} \emph{closeness}.
In total, $7.85\times10^{6}$ features are present in the dataset;
the lifetime and \emph{average} \emph{closeness} quantities are used
to filter features in the remainder of this paper (\S\S\ref{subsec:Magnetic-Feature-Selection}
\& \ref{subsec:Latitudinal-Flows}).

\subsection{\label{subsec:Longitudinal-Flows}Longitudinal Flow Measurement}

\subsubsection{\label{subsec:The-Synodic-Sidereal-Correction}The Synodic-Sidereal
Correction}

After tracking, the features' pixel locations stored in the \emph{feature}
\emph{table} are converted heliographic coordinates (central meridian
distance $l$ and latitude $b$) using a perspective transformation
for each image. For each feature, I used equations (1) and (2) of
\citet{Sudar2015} to estimate $\omega_{syn}$ and $\omega_{mer}$.
Those equations are the ordinary least squares estimates of\textbf{
}$\omega_{syn}$ and $\omega_{mer}$ using a simple regression model.
To convert the synodic rotation rate $\omega_{syn}$ to the sidereal
rotation rate $\omega_{sid}$ I also used equation (7) of \citet{Skokic2014}:
\begin{equation}
\omega_{sid}=\omega_{syn}+\frac{\bar{\omega}_{Earth}}{r^{2}}\frac{\cos^{2}\Psi}{\cos i}.\label{eq:skokic7}
\end{equation}
In that equation, $\bar{\omega}_{Earth}=0.9856\text{ deg}\textrm{ d}^{-1}$
is Earth's yearly-averaged angular speed, $r$ is the dimensionless
Earth-Sun distance expressed in AU, $i$ is the inclination of the
solar equator to the ecliptic, and $\Psi$ is the angle between the
pole of the ecliptic and the solar rotation axis orthographically
projected onto the solar disk. The first fraction $\frac{\bar{\omega}_{Earth}}{r^{2}}$
gives Earth's instantaneous angular speed $\omega_{Earth}$. In the
second fraction, $\Psi$ is calculated from
\begin{equation}
\tan\Psi=\tan i\cos(\lambda_{0}-\Omega),\label{eq:orthographic-projection}
\end{equation}
where $\lambda_{0}$ is the apparent longitude of the Sun with respect
to the true equinox on the date of observation, and $\Omega$ is the
longitude of the ascending node of the solar equator on the ecliptic.
The quantities $r$ and $\lambda_{0}$ can be retrieved from the the
JPL Horizons ephemerides\footnote{\url{http://ssd.jpl.nasa.gov/?horizons}}
by selecting the Observer ephemeris type and including the ``Observer
range \& range-rate'' and the ``Observer ecliptic lon. \& lat.''
output quantities, respectively\emph{.} For $i$ I use 7.25\degree,
and for $\Omega=73\degree40'+50.25"(t-1850.0)$, where $t$ is
the time expressed in fractional years \citep{Rosa1995SoPh}. As an
example, for 2011-Feb-01 00:00:00 UTC, $r=0.9853$, thus $\lambda_{0}=311.76\degree$,
and $\Omega=75.92\degree$. Thus $\Psi=-4.0852\degree$,
$\cos^{2}(\Psi)/\cos(i)=1.002942$. $\omega_{Earth}=1.01523\text{ deg}\textrm{ d}^{-1}$
and thus the difference between the calculated sidereal and measured
synodic rotation rates is $1.018215\text{deg}\,\textrm{d}^{-1}$.
As a check, I note that this is faster than average $(\bar{\omega}_{Earth})$
because Earth is near perihelion in February. For a feature first
detected at this time with $\omega_{syn}=12.170066\text{ deg}\,\text{d}^{-1}$,
I thus compute $\omega_{sid}$ to be $13.188281\text{ deg}\,\text{d}^{-1}$.

\subsubsection{\label{subsec:Magnetic-Feature-Selection}Magnetic Feature Selection
and Filtering}

Magnetic features are subject to motion by a bevy of plasma flows.
At the 12~min data cadence used here, granulation flows changing
on time scales of 5\textendash 10 minutes act to buffet the features,
while super- and (perhaps) meso-granular flows can act to produce
sustained feature motions over $\sim$24~h and 0.5-4~h respectively.
In addition to these flows, interactions among features can cause
apparent motion of the feature centroids. During a merging of two
like-signed features, the feature with the larger flux retains its
identification number (assured by the dual flux-weighted overlap criterion
of the feature tracking's association step), so the addition of flux
from the smaller feature can cause the flux-weighted centroid location
of the larger feature to jump in the direction of the smaller feature
after the merge. Therefore understanding and controlling for the effect
of feature interactions is an important step in separating systematic
from real effects.

I use two parameters to control the number of features in the ensemble
used for measuring the rotational flow $\omega_{sid}$: the lifetime,
and the \emph{average} \emph{closeness} of each feature. The rotational
measurement is largely insensitive to the choice of parameters, but
there are some combinations that provide smaller error bars and smoother
average rotational profiles than others, as shown below. I created
ensembles of features that had lifetimes greater than certain thresholds,
and average closeness less than certain thresholds. In general, if
the thresholds are made excessively strict then insufficient numbers
of features are available for averaging and even the fact that the
Sun rotates differentially can not be conclusively determined. If
the thresholds are made too lenient other processes such as the aforementioned
merging substantially affect the measurement.

\begin{figure}
\begin{centering}
\includegraphics{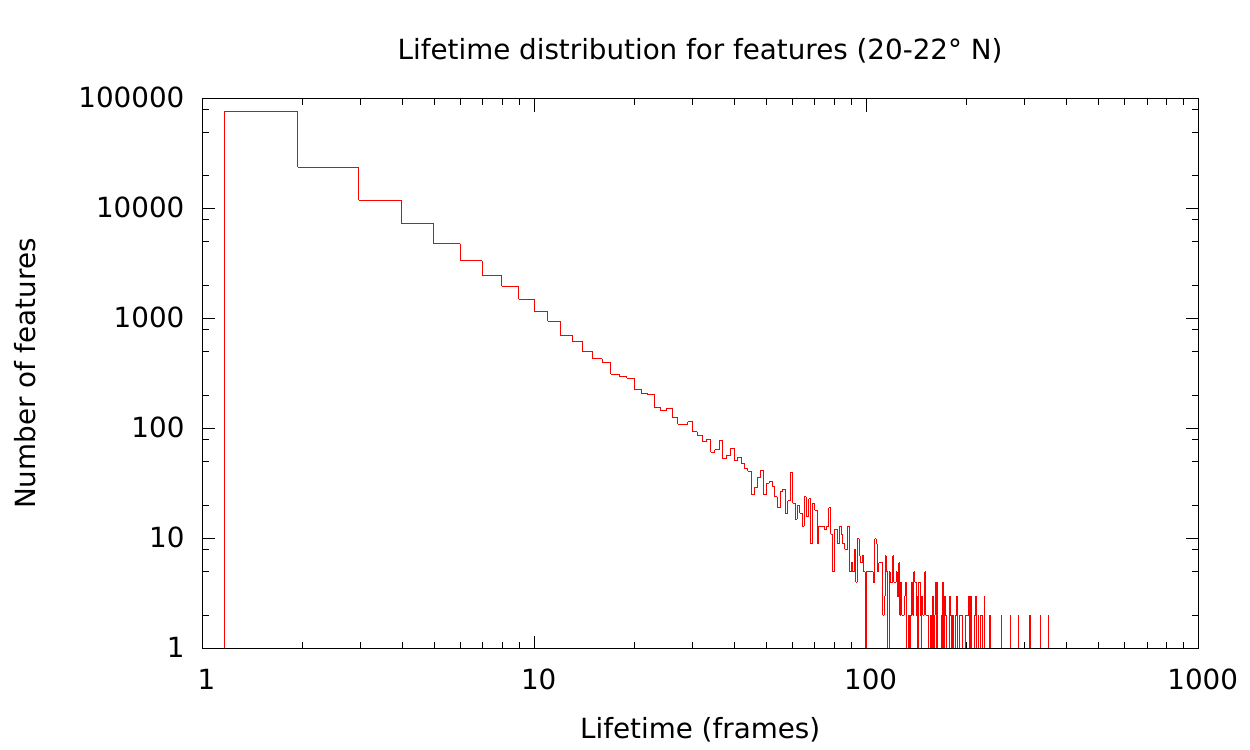}
\par\end{centering}
\caption{\label{fig:Lifetime-distribution}Lifetime distribution of features
in the data set, in a single 20\textendash 22\degree~N latitudinal
bin, with a 1-frame binning. No correction has been made for the data
gaps, which would result in a small shift to the right of some of
the features. Note that both axes are logarithmically scaled.}
\end{figure}

The feature lifetime distribution is steep and dominated by features
with short lifetimes (Figure~\ref{fig:Lifetime-distribution}), and
fragmentation and merging are the dominant methods
of feature birth and death \citep{2013ApJ...774..127L}.
 Because of this, an ensemble with a short lifetime threshold will
include many features that are born and die by these methods. The
recorded birth and death frame of each feature are the frames in which
the feature is first or last recognized, and not instead the frame
in which the interaction that resulted in the birth or death was first
determined to begin. Since the interactions may take multiple frames
to complete, the linear fit of the feature's position over time (equations
(1) \& (2) of \citet{Sudar2015}) can be strongly affected by the
birth and death interactions if the lifetime threshold is chosen to
be too low. In addition, making the average closeness threshold less
strict allows for more feature-feature interactions, and thus the
position of the flux-weighted centroid of a feature experiencing an
interaction undergoing merging will have large jumps that are not
due to the flow but rather to these interactions. The surface density
of coronal bright points ($\sim$100 on the disk at any one time)
is two orders of magnitude lower than the average surface density
of magnetic features, so the average closeness threshold should be
set so as to exclude those features that are always nearby (and presumably
interacting with) other features. 

\begin{sidewaysfigure}
\includegraphics[width=0.32\textwidth]{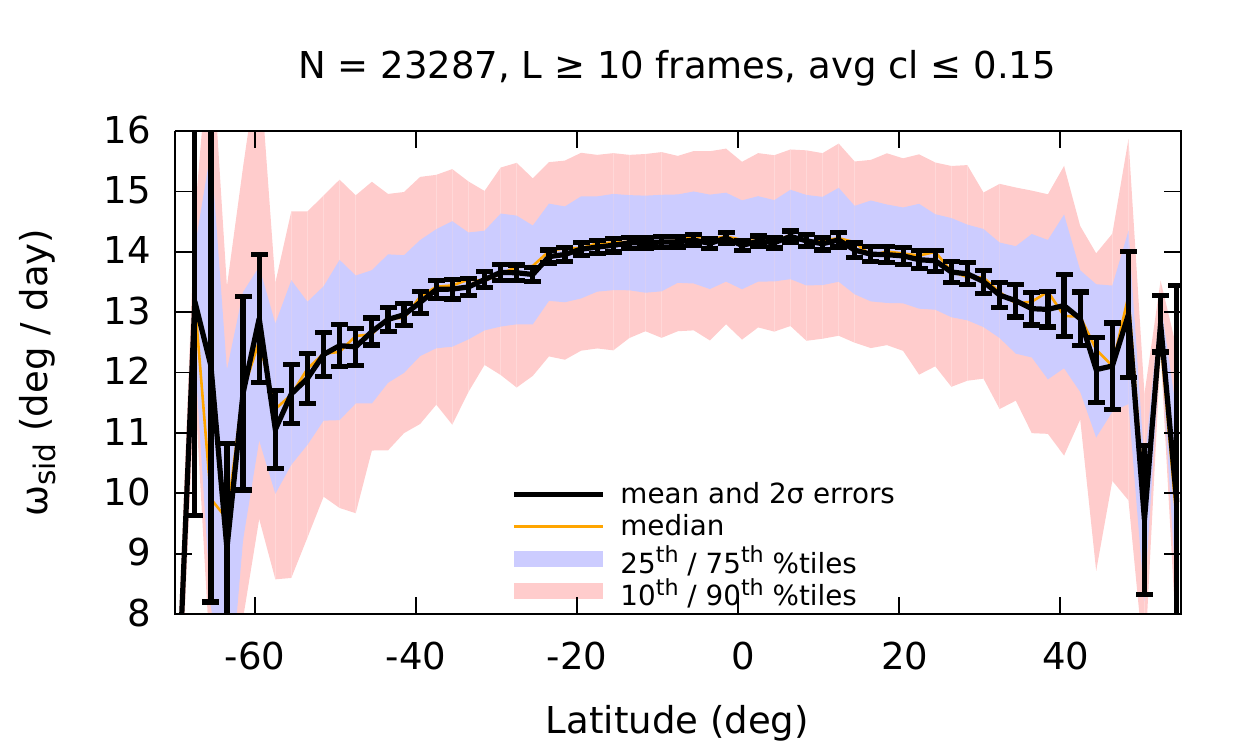}\includegraphics[width=0.32\textwidth]{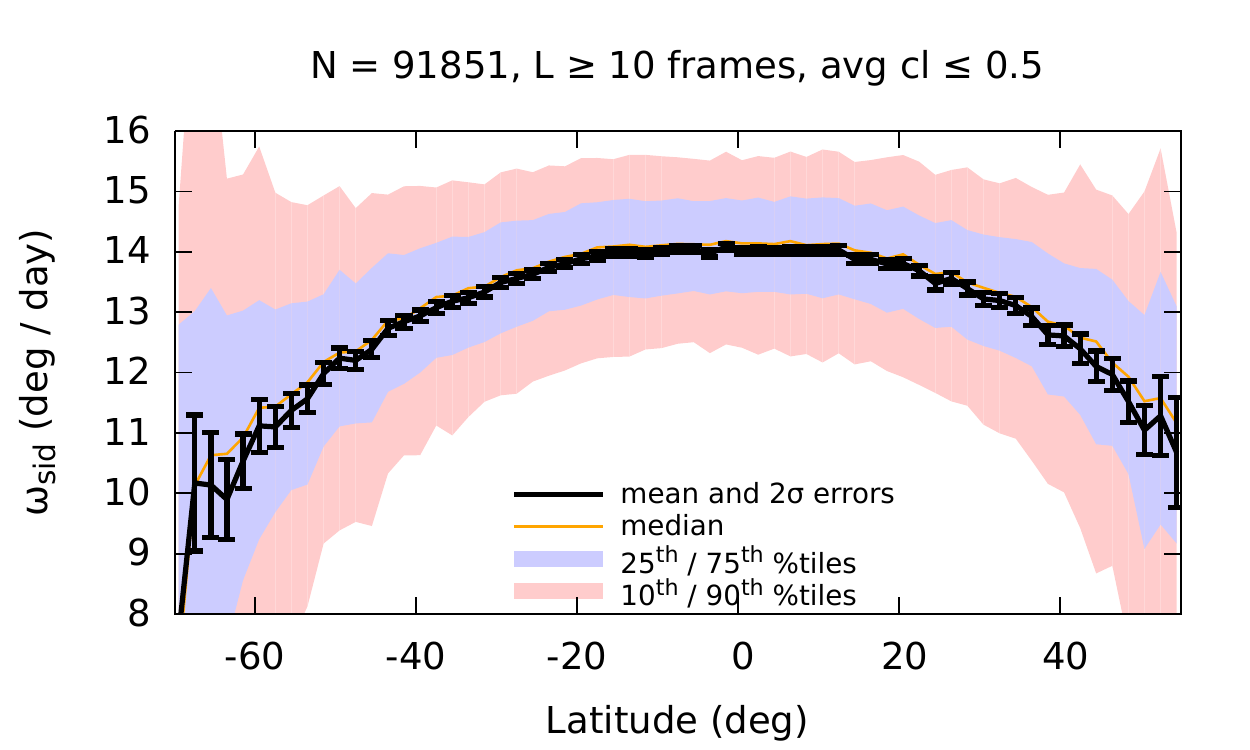}\includegraphics[width=0.32\textwidth]{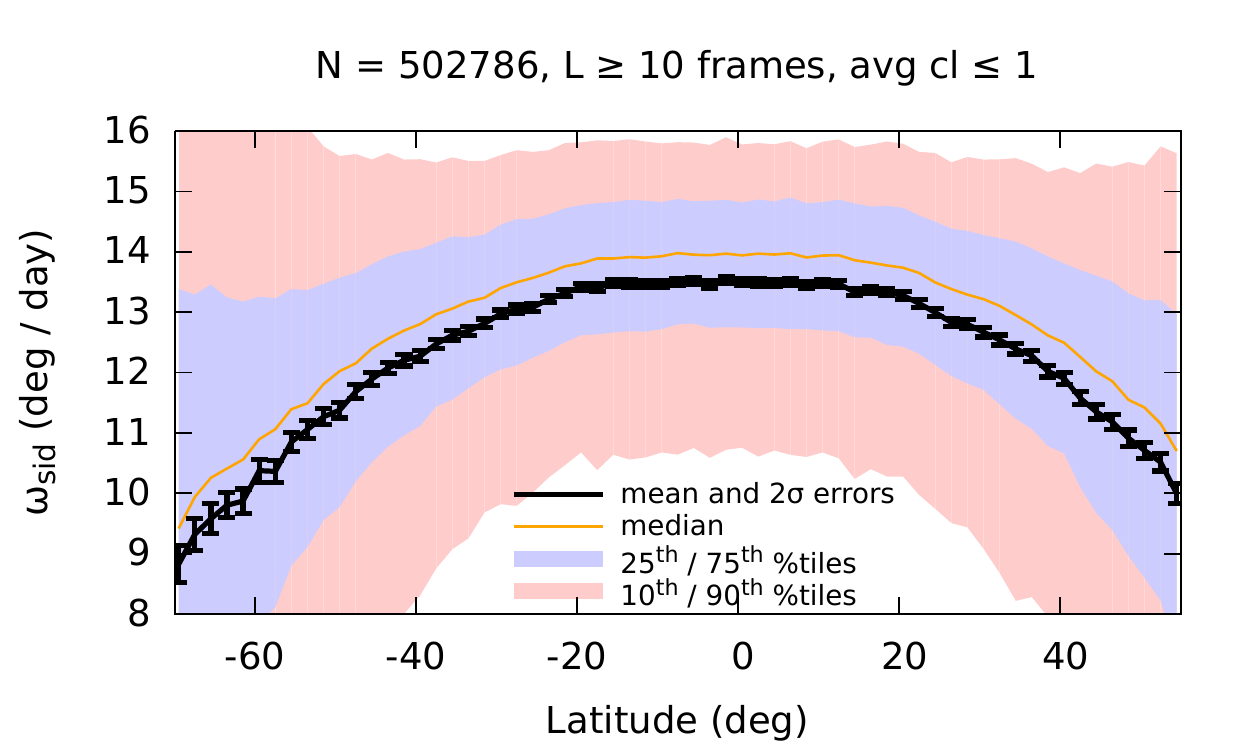}

\includegraphics[width=0.32\textwidth]{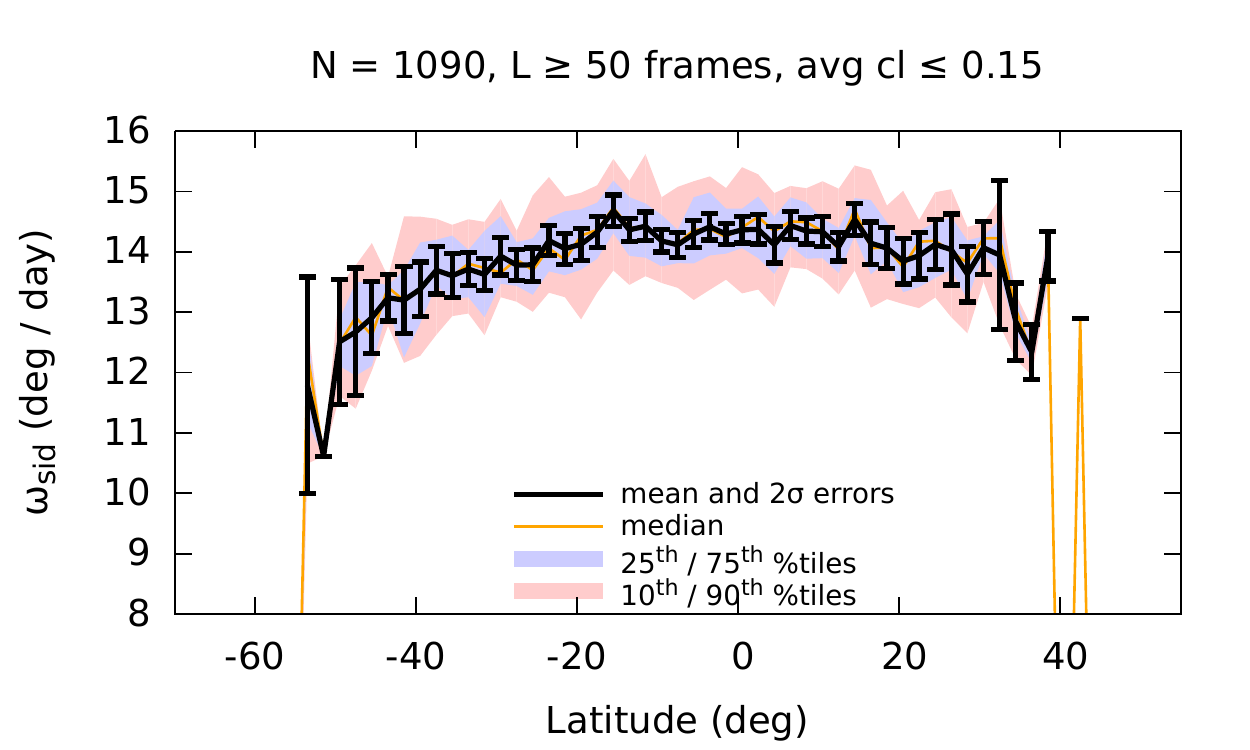}\includegraphics[width=0.32\textwidth]{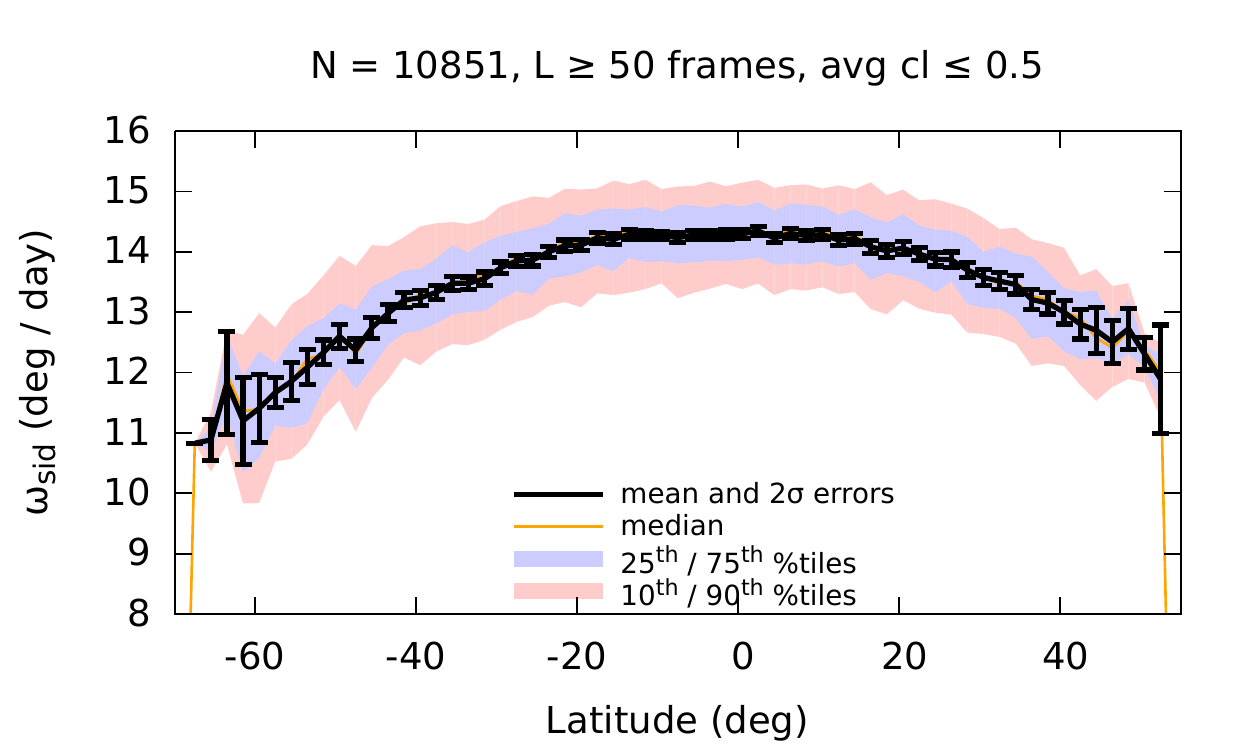}\includegraphics[bb=0bp 0bp 587bp 368bp,clip,width=0.31\textwidth]{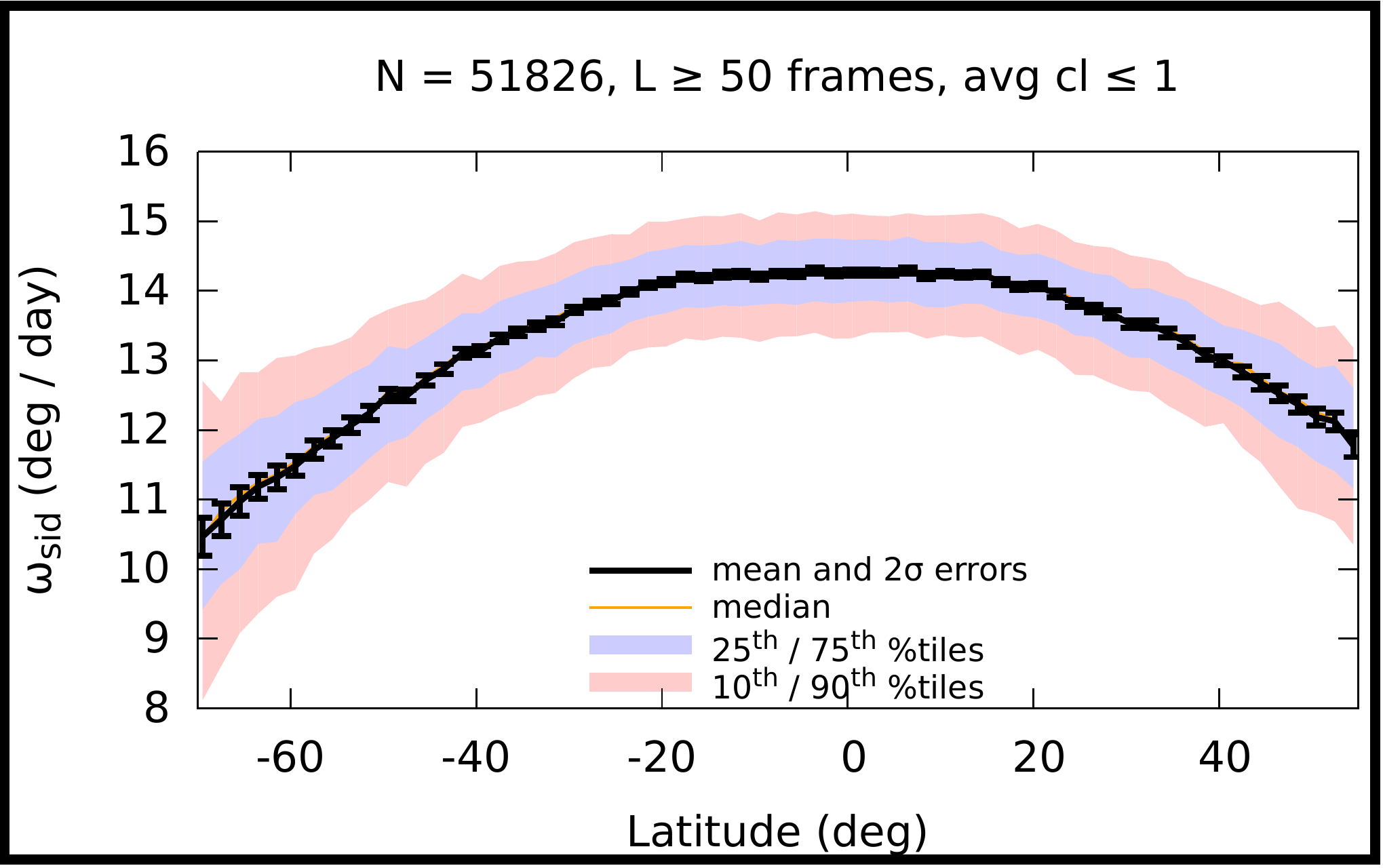}

\includegraphics[width=0.32\textwidth]{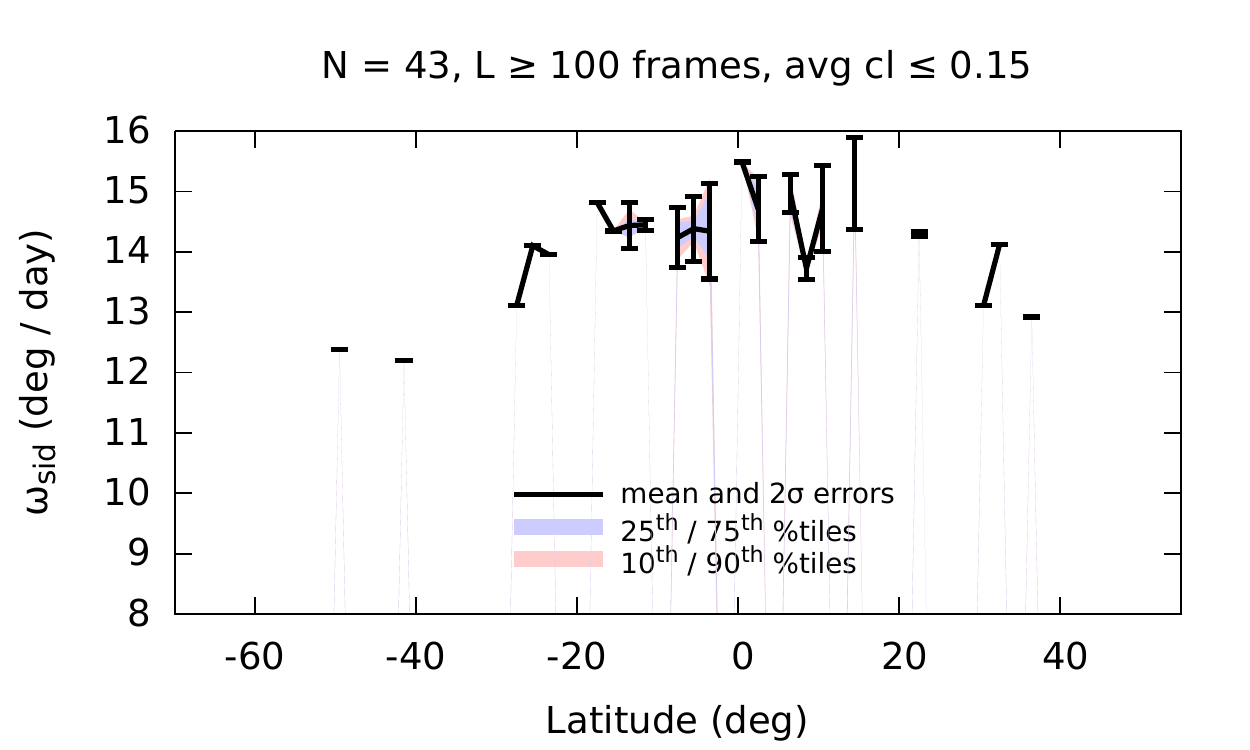}\includegraphics[width=0.32\textwidth]{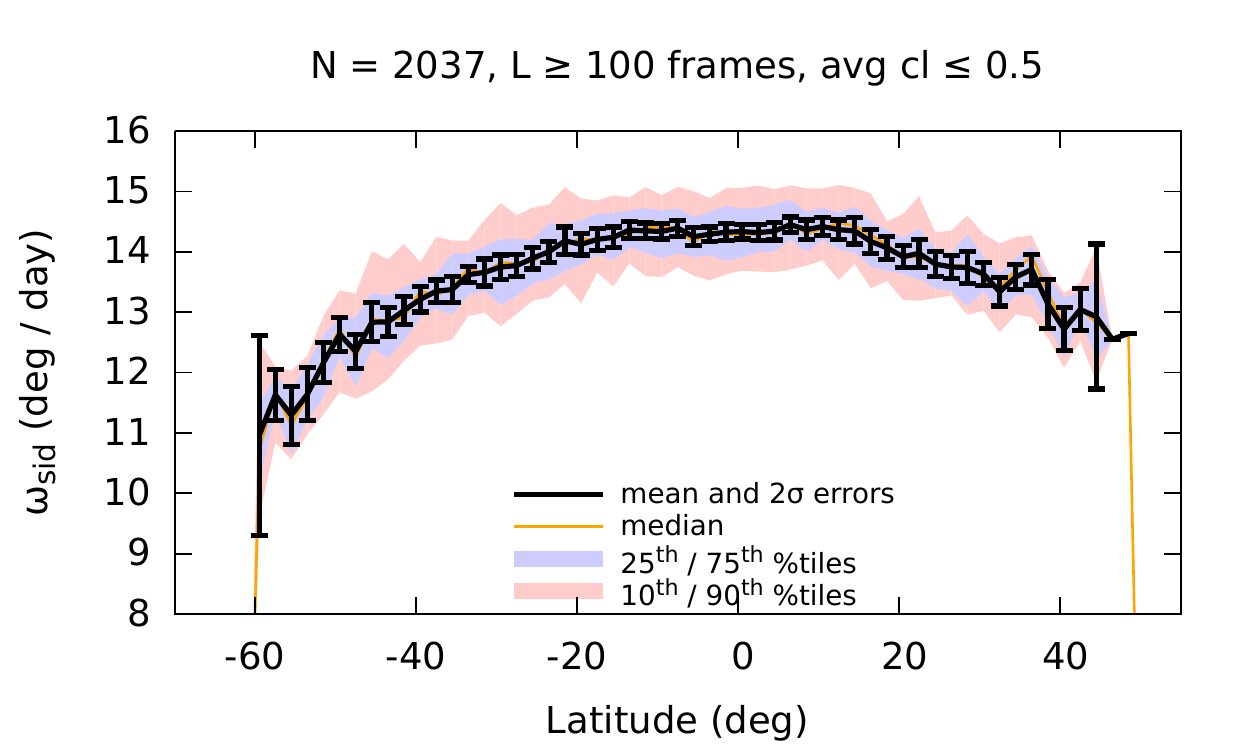}\includegraphics[width=0.32\textwidth]{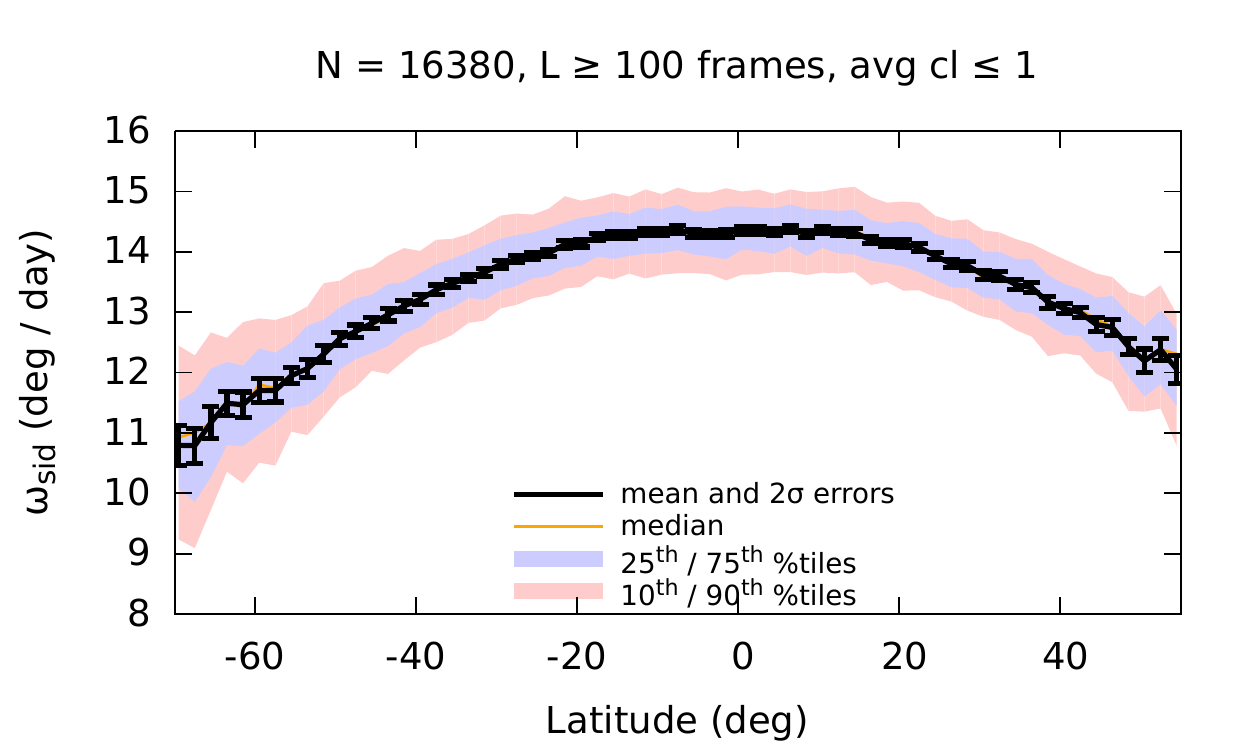}

\caption{\label{fig:rotational-motion-for}Rotational motion for lifetime (rows;
top to bottom: 10, 50, 100) and average closeness parameters (columns;
left to right: 0.15, 0.50, 1.00). The latitudinal bin size in all
plots is 2 degrees. The combination of the lifetime and average closeness
parameters strongly affects the smoothness of the mean profiles and
the width and symmetry of the distributions in the latitudinal bins.
I use the boxed plot for the rotational profile analysis in the remainder
of the paper.}
\end{sidewaysfigure}

Figure~\ref{fig:rotational-motion-for}
 shows rotational flow measurements for all combinations of lifetime
$\ge$ (10, 50, 100) frames and average closeness $\le$ (0.15, 0.5,
1.0). For lifetime $\ge$ 100 and average closeness $\le$ 0.15 (lower-left
panel), I am selecting features that live $\ge$ 20 hours and have
a neighboring feature less than 15\% of the time. Since even the supergranular
network turns over on timescales of $\sim$24~h, and the average
closeness threshold excludes feature-dense regions such as active
regions, it is not surprising that only 43 features satisfy these
criteria, making even a measurement of the rotational motion unreliable.
On the opposite end of the range (upper right panel), I find that
changing the selection thresholds so as to include the largest number
of features is not necessarily desirable. For lifetime $\ge$10~frames
and average closeness $\le1$ (i.e., no restriction), over $5\times10^{5}$
features are found. Even though the 2$\sigma$ error bars are very
small, the median of each bin is much higher from the mean than those
error bars\textendash a large number of low velocity features depress
the mean from the median in each bin\textendash and the distributions
are much wider than would be expected on physical grounds. That the
low-velocity tail so substantially affects the mean suggests that
suggests that additional filtering of features is necessary in order
to bring the mean and median into closer agreement and reduce the
outsize effect of the tails of the distribution in each latitudinal
bin.

\citet{Sudar2015} discarded any bright points with sidereal speeds
$\le8$ or $\ge19\mbox{ deg}\text{ d}^{-1}$. I can explicitly discard
those features, or achieve the same results (while also potentially
discarding a small number of desirable features) by using a lifetime
threshold of at least 28~frames (5.6~h). Over the range of lifetime
thresholds between 30 and 100~frames the mean of the sidereal rotation
between 20 and 22\degree latitude differs only by $0.2\text{ deg}\text{ d}^{-1}$,
the median only by $0.13\text{ deg}\text{ d}^{-1}$. The difference
between the two, which is indicative of asymmetries in the velocity
distributions, is less than $0.1\text{ deg}\text{ d}^{-1}$ over this
range (Table~\ref{tab:w_sid-Mean-Median-vs-Lifetime}). For the remainder
of the rotational flow results, I adopt a minimum lifetime of 50~frames
as a reasonable compromise between rejecting outliers and having a
large enough sample size in each bin. It is at this point in Table~\ref{tab:w_sid-Mean-Median-vs-Lifetime}
that the difference between the mean and median in that particular
latitudinal bin first reaches a local minimum.
\begin{table}
\begin{centering}
\caption{\label{tab:w_sid-Mean-Median-vs-Lifetime}Mean and median sidereal
rotation velocities, and their difference, and the number of features,
for features with $20\degree\le b\le22\degree$ and given
minimum lifetime thresholds.}
\begin{tabular}{ccccc}
\hline 
Min Lifetime (frames) & Mean $(\degree\text{d}^{-1})$ & Median $(\degree\text{d}^{-1})$ & Difference $(\degree\text{d}^{-1})$ & N\tabularnewline
\hline 
\hline 
10 & 13.263 & 13.716 & 0.453 & 9426\tabularnewline
20 & 13.812 & 13.936 & 0.124 & 3768\tabularnewline
30 & 13.912 & 13.991 & 0.080 & 2211\tabularnewline
40 & 13.974 & 14.031 & 0.058 & 1496\tabularnewline
50 & 14.031 & 14.059 & 0.028 & 1101\tabularnewline
60 & 14.040 & 14.074 & 0.035 & 829\tabularnewline
70 & 14.067 & 14.098 & 0.031 & 650\tabularnewline
80 & 14.077 & 14.099 & 0.022 & 524\tabularnewline
90 & 14.089 & 14.099 & 0.010 & 424\tabularnewline
100 & 14.110 & 14.127 & 0.017 & 365\tabularnewline
\hline 
\end{tabular}
\par\end{centering}
\end{table}

\begin{figure}
\includegraphics[width=1\textwidth]{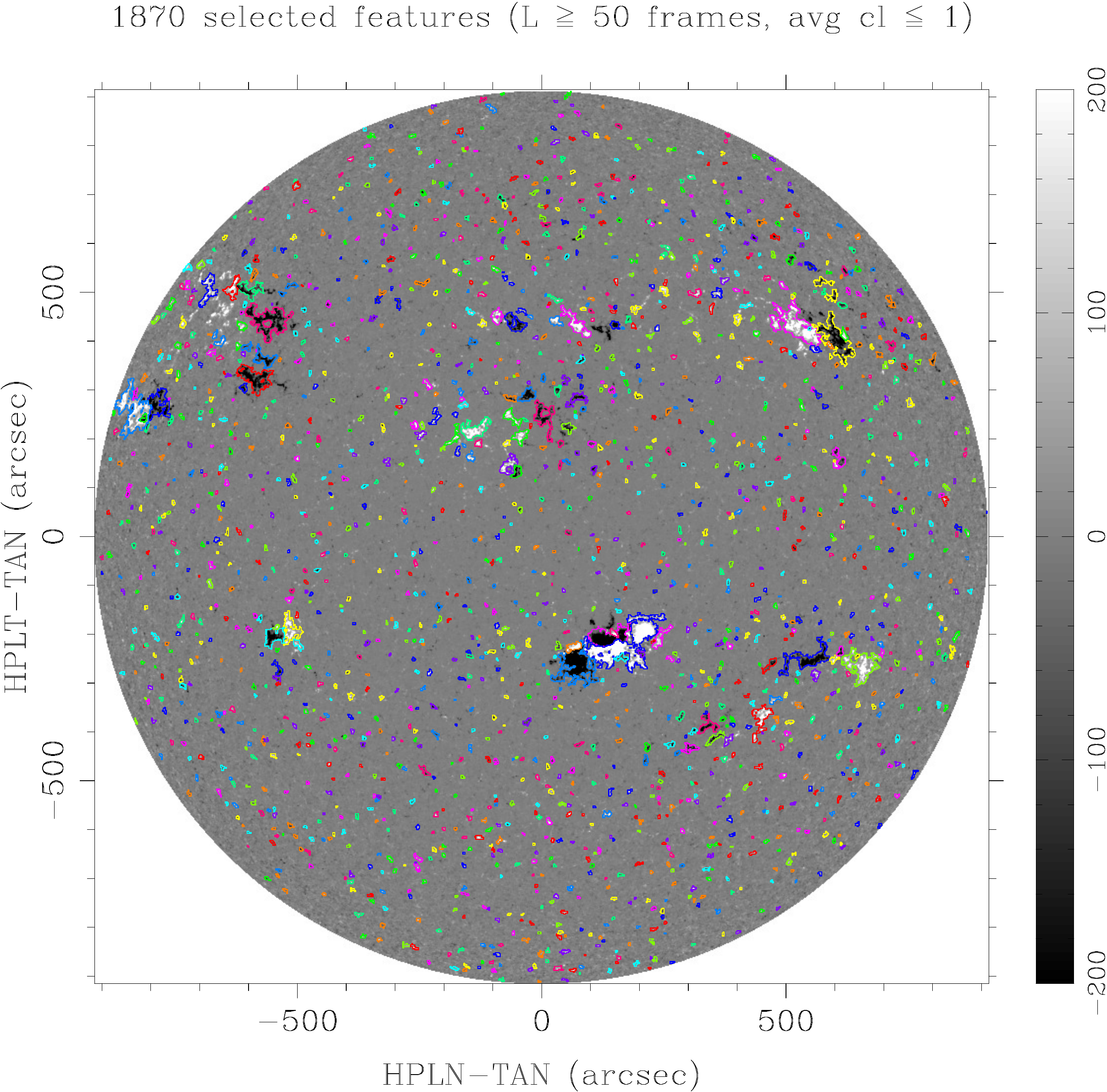}

\caption{\label{fig:selected-features}Magnetogram from 2011-02-14 17:36:45
TAI with the 1870 features that are identified by SWAMIS in that frame
and that satisfy my parameter thresholds (lifetime \ensuremath{\ge}
50~frames, average closeness \ensuremath{\le} 1.0). The colors of
the feature outlines are for clarity only; similarities or differences
among them carry no scientific meaning. Pixels where $\alpha\ge70\degree$
have been cropped out, as described in the \S~\ref{sec:Data-Observations}
text. The magnetogram grayscale saturates as $\pm200$~G.}

\end{figure}

For the 50-frame lifetime threshold, I also produced plots like those
in Figure~\ref{fig:rotational-motion-for} for average closeness
thresholds of 0.8 and 0.9. Even though those thresholds excluded active
regions but the threshold of 1.0 includes active regions, I find no
substantial difference in the rotational profile. For a threshold
of 1.0, the rotation profile is essentially unchanged at the active
latitudes, and is smoother at higher absolute latitudes (due to larger
number of features). Thus for the remainder of the rotational flow
analysis, I adopt the parameters of (lifetime \ensuremath{\ge} 50~frames,
average closeness \ensuremath{\le} 1.0). Figure~\ref{fig:selected-features}
shows the features selected by these parameters for a single magnetogram.

\subsection{\label{subsec:Latitudinal-Flows}Latitudinal Flow Measurement}

\begin{sidewaysfigure}
\includegraphics[width=0.32\textwidth]{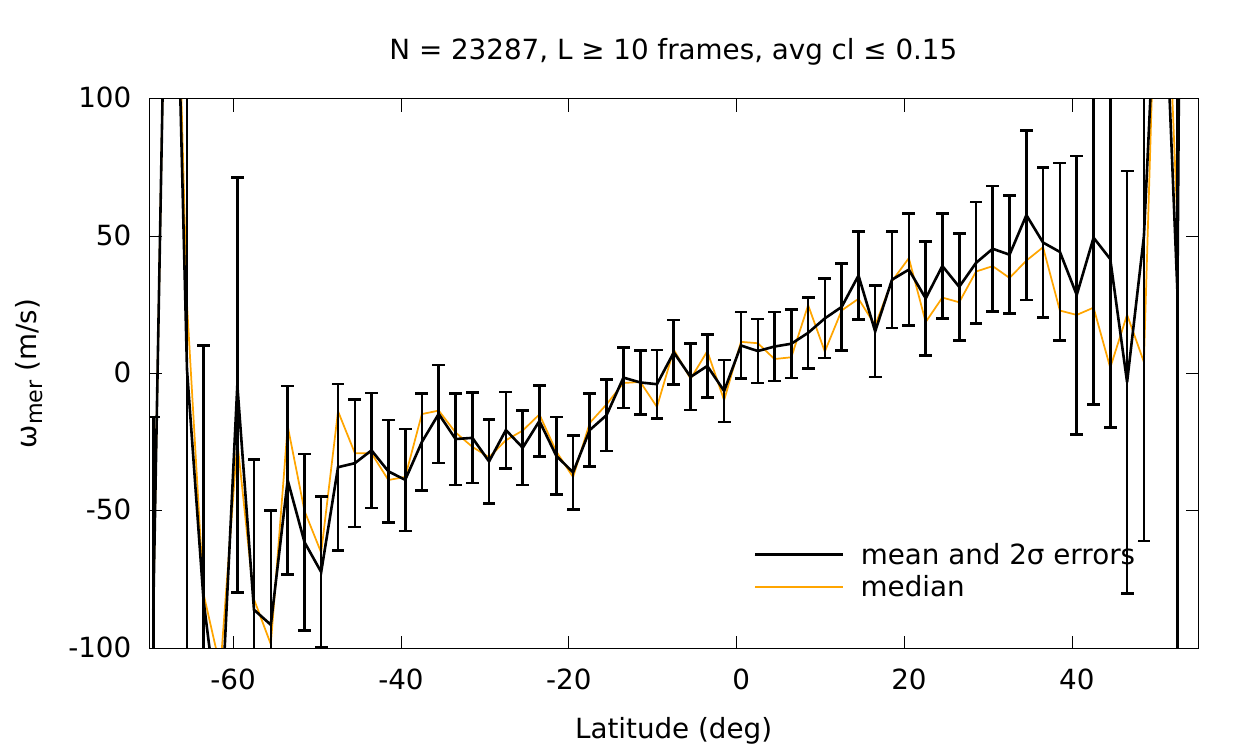}\includegraphics[width=0.32\textwidth]{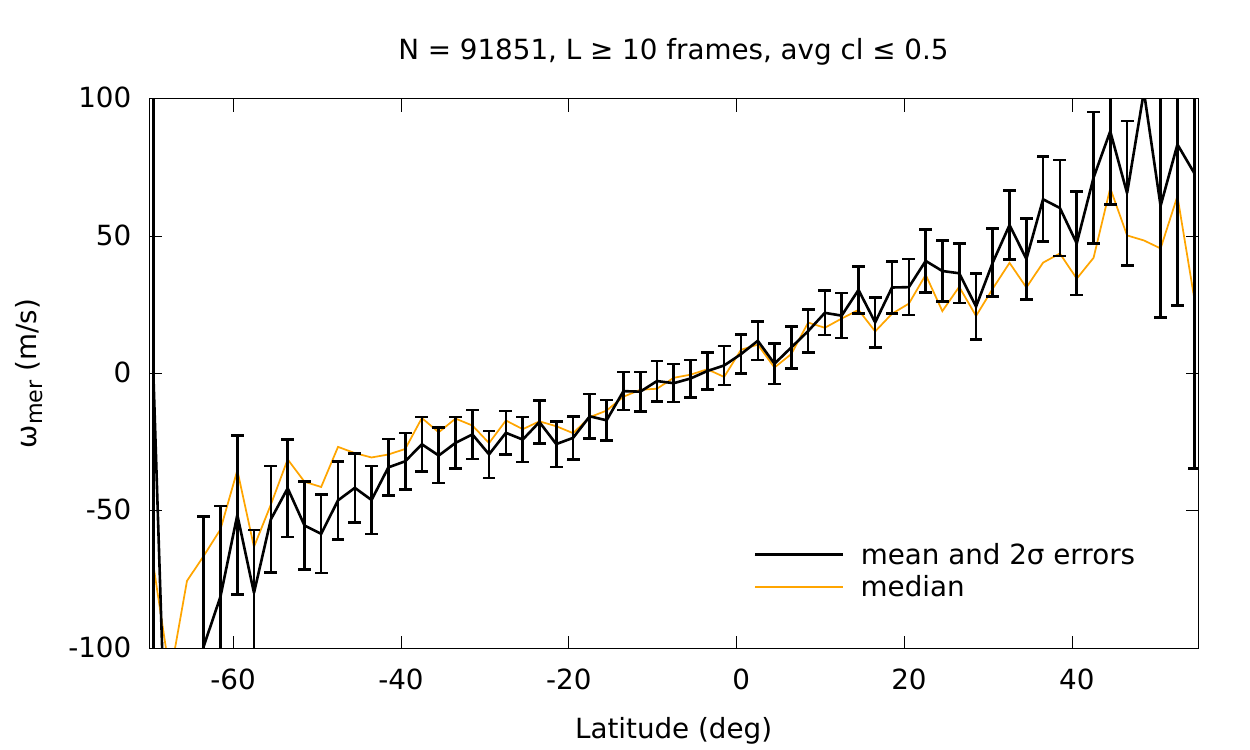}\includegraphics[width=0.32\textwidth]{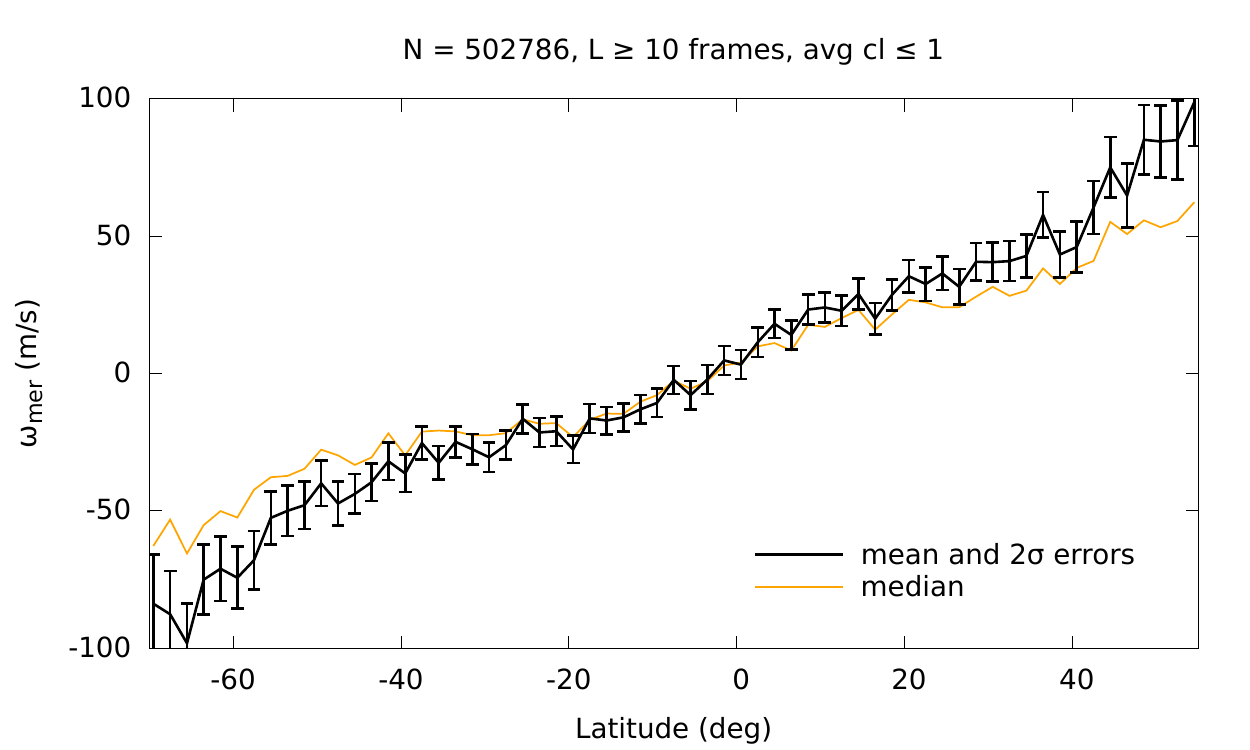}

\includegraphics[width=0.32\textwidth]{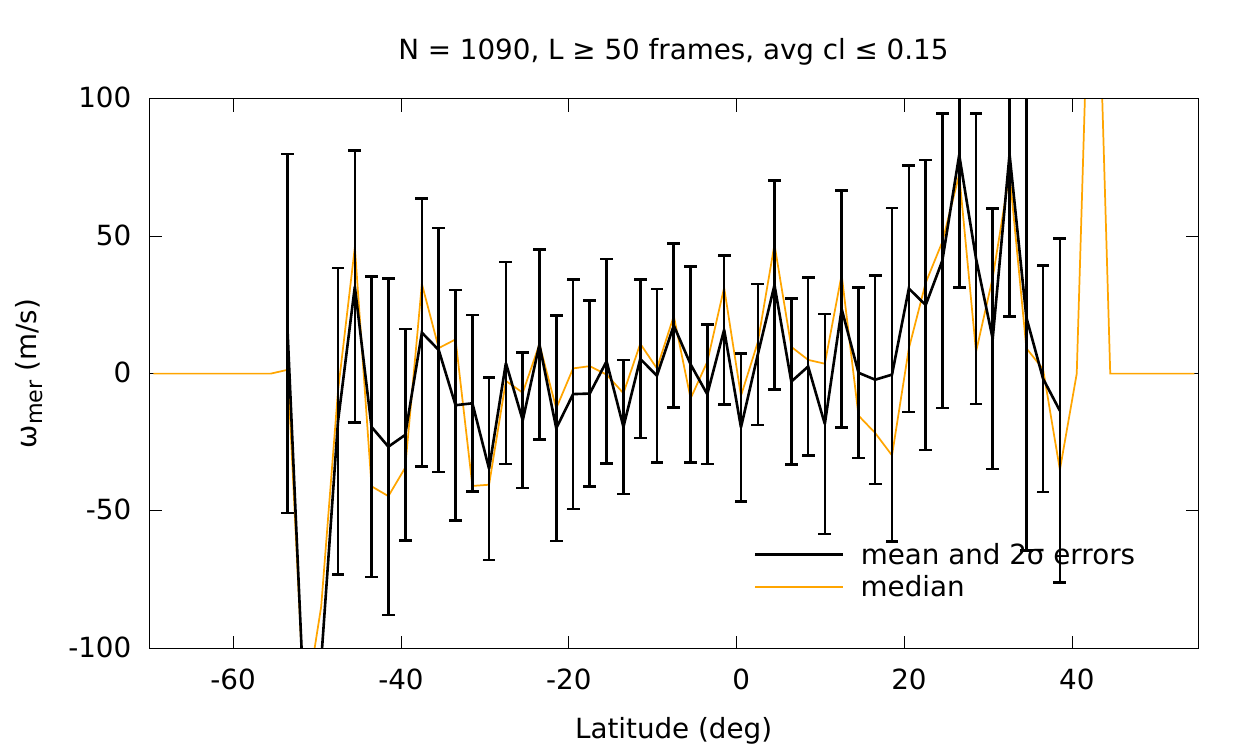}\includegraphics[width=0.32\textwidth]{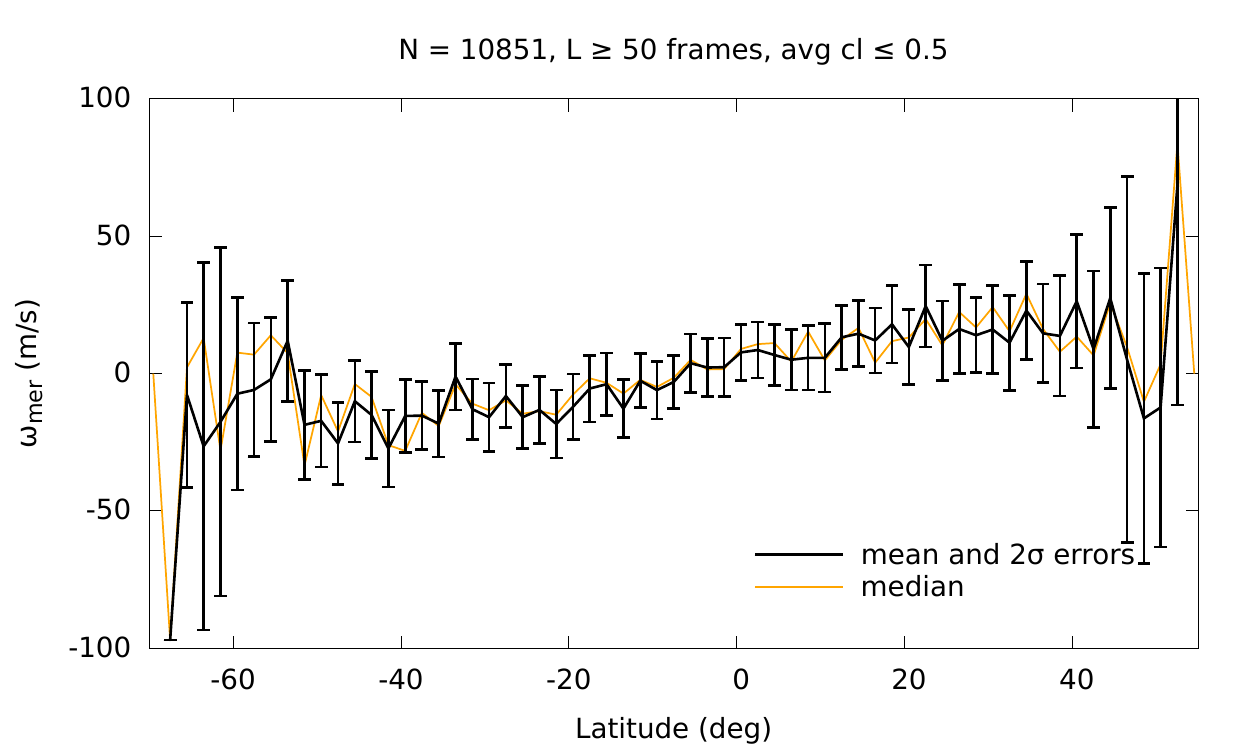}\includegraphics[width=0.32\textwidth]{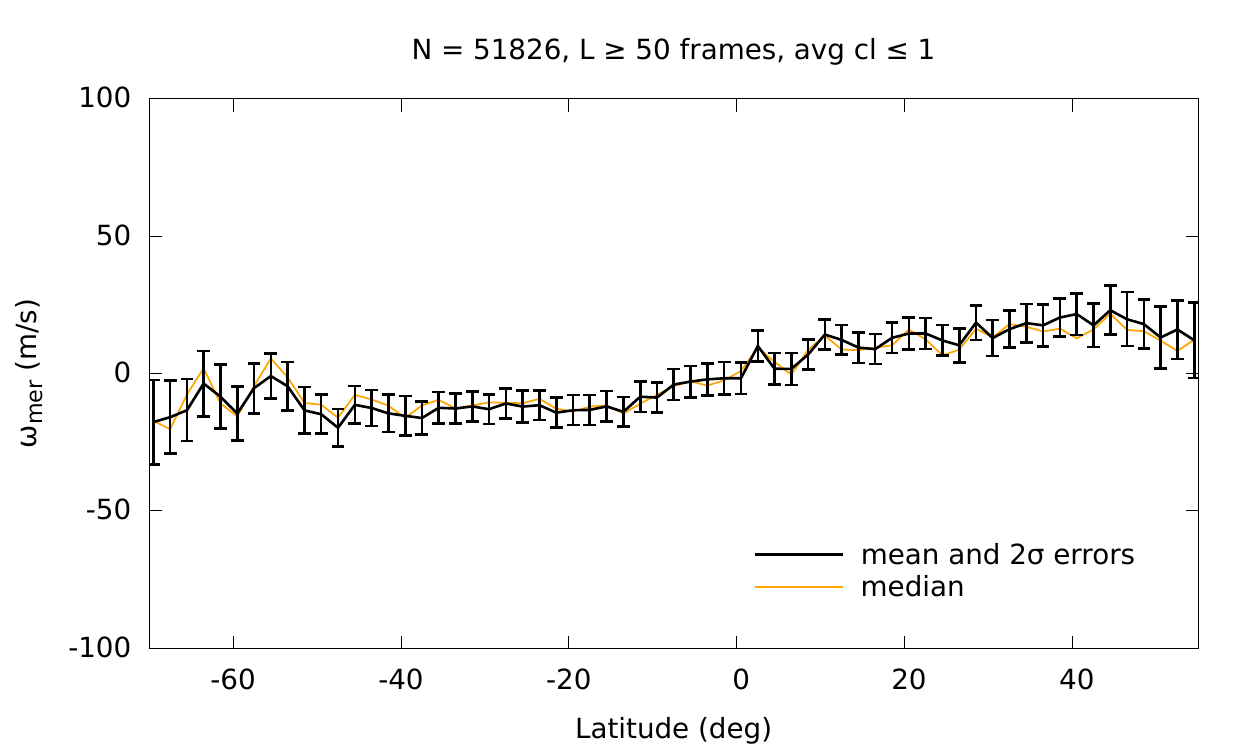}

\includegraphics[width=0.32\textwidth]{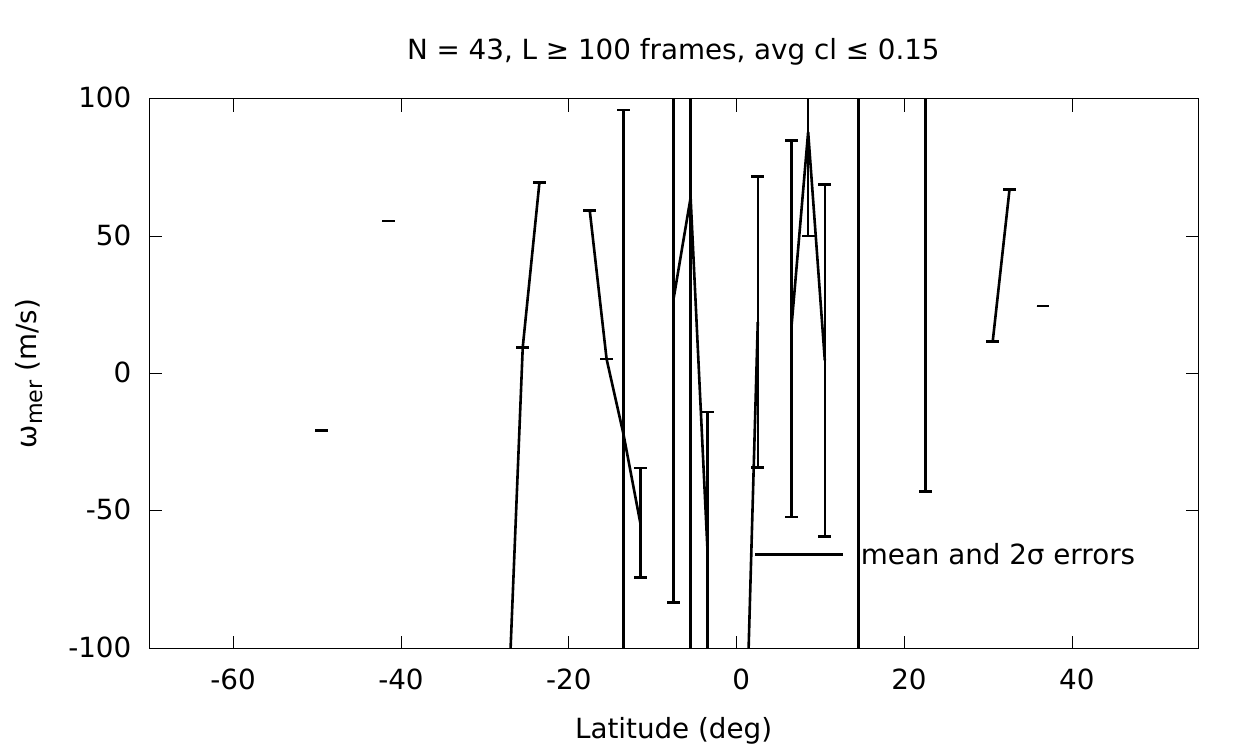}\includegraphics[width=0.32\textwidth]{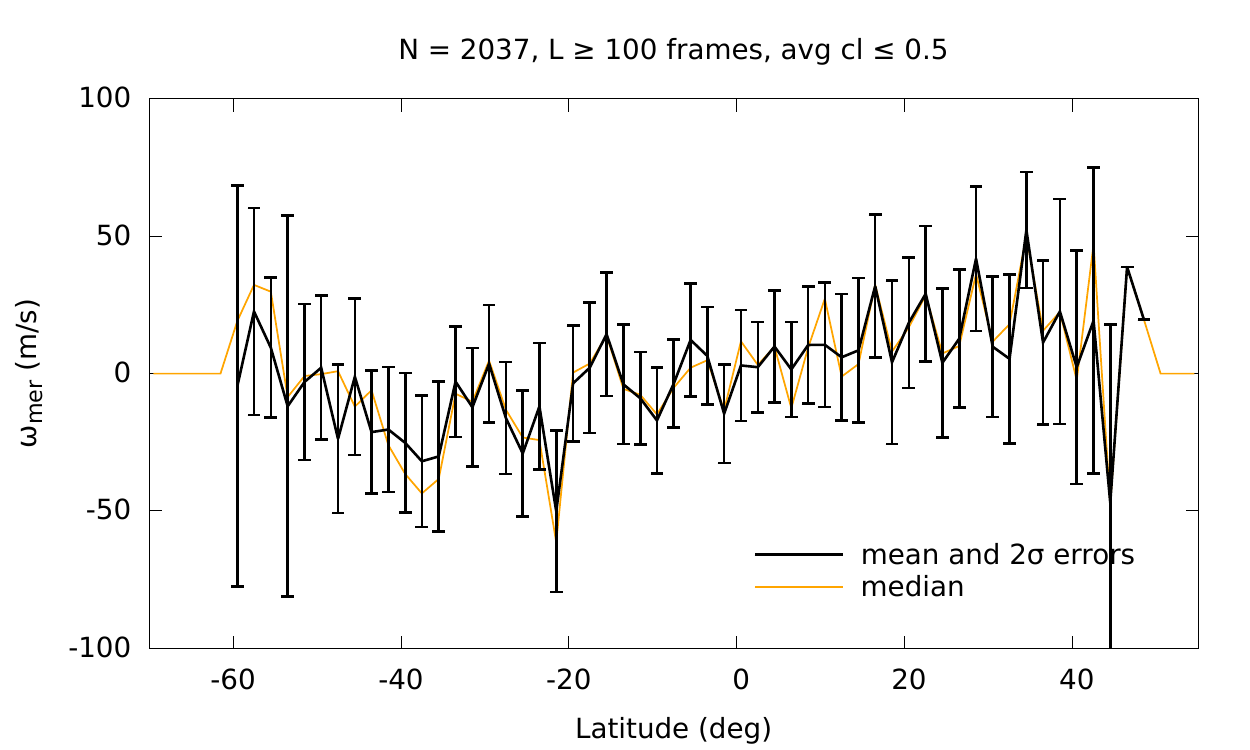}\includegraphics[width=0.32\textwidth]{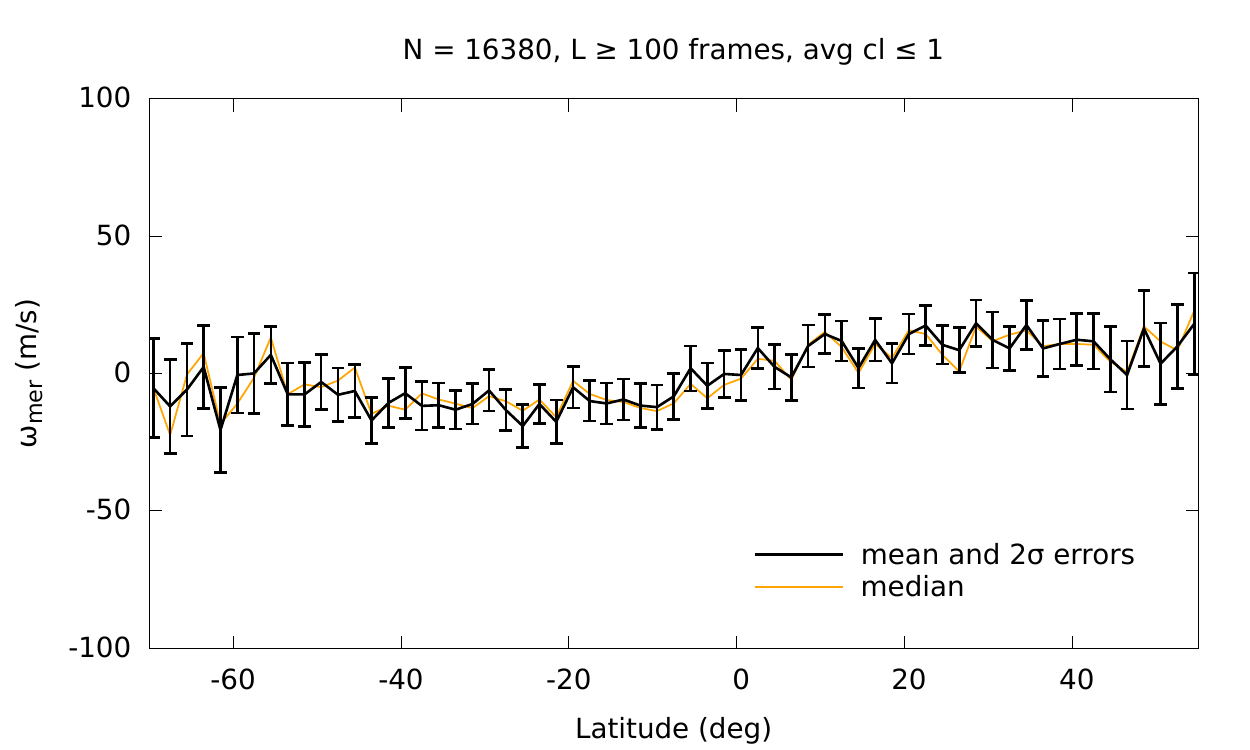}

\caption{\label{fig:meridional-motion-for}Similar to Figure~\ref{fig:rotational-motion-for},
but for the features' meridional speed given lifetime thresholds (rows;
top to bottom: 10, 50, 100) and average closeness parameters (columns;
left to right: 0.15, 0.50, 1.00). Again, the combination of the lifetime
and average closeness parameters strongly affects the smoothness and
apparent functional form of the mean profiles and the width and symmetry
of the distributions in the latitudinal bins.}
\end{sidewaysfigure}

Because the meridional flow is 2\textendash 3 orders of magnitude
slower than the rotational flow, by all methods used to date it is
more difficult to measure and has larger uncertainties than rotational
flow measurements. Figure~\ref{fig:meridional-motion-for}
 shows the meridional flow measurement for the same combination of
lifetime and average closeness thresholds as in Figure~\ref{fig:rotational-motion-for}.
For the measurement of the meridional velocities, I note again that
several combinations of thresholds results in curves that are too
noisy to be meaningful: all measurements for which average closeness
$\le0.15$, (100, 0.5), and (50, 0.5). Combinations (10, 0.5) and
(10, 1) are noticeably different from (50,1) \& (100, 1): the former
two have very high speeds at high latitudes, with no evidence of the
mid-latitude peak in the meridional speed that is seen to some degree
in the latter two observations. That peak seems to suggest a functional
form of $\omega_{mer}\propto2\sin\theta\cos\theta$ \citep{Hathaway2010_MeridionalFlow_SolarCycle},
though I must understand the difference between the two trends and
not simply choose the one that matches other observations. Considering
the difference between (50, 1) and (10,1), the latter includes 10
times as many features, and the excess features all have lifetimes
between 10 and 49 frames. As for the rotational flow, the mean and
median of each bin do not agree, particularly in the higher latitudes,
similar to how the mean and median rotational flow did not agree for
some parameter combinations. I again examine the difference between
the mean and median in the bins. Because of the noisier plots I calculate
the mean and median of each 2-degree-wide bin in the range 20\degree\textendash 30\degree~N.
Table~\ref{tab:w_mer-Mean-and-median-vs-Lifetime} shows the mean
of the bin means, the mean of the bin medians, and the mean of the
absolute value of their differences, as a function of lifetime threshold.
\begin{table}
\caption{\label{tab:w_mer-Mean-and-median-vs-Lifetime}Mean and median meridional
flow velocities, and their difference, and the number of features,
for features with $20\degree\le b\le30\degree$ divided
into 2\degree bins and given minimum lifetime thresholds.}

\centering{}%
\begin{tabular}{ccccc}
\hline 
Min Lifetime (frames) & <Mean> $(\text{m/s})$ & <Median> $(\text{m/s})$ & <|Difference|> $(\text{m/s})$ & N\tabularnewline
\hline 
\hline 
10 & 35.582 & 25.911 & 9.671 & 44328\tabularnewline
20 & 23.472 & 18.408 & 5.065 & 17350\tabularnewline
30 & 18.243 & 15.060 & 3.183 & 10097\tabularnewline
40 & 15.346 & 12.449 & 2.969 & 6813\tabularnewline
50 & 13.825 & 11.781 & 2.611 & 5044\tabularnewline
60 & 14.094 & 12.069 & 2.745 & 3813\tabularnewline
70 & 13.233 & 11.731 & 2.499 & 2996\tabularnewline
80 & 13.162 & 11.725 & 2.497 & 2445\tabularnewline
90 & 13.246 & 11.845 & 2.259 & 2002\tabularnewline
100 & 14.151 & 12.049 & 2.840 & 1658\tabularnewline
\hline 
\end{tabular}
\end{table}
 Again, the difference between the mean of the bin means and the mean
of the bin medians has its first local minima at a lifetime threshold
of 50~frames, as for the lifetime distributions.

\begin{sidewaysfigure}
\begin{minipage}[c][1\totalheight][t]{0.5\columnwidth}%
\vspace{0pt}
(a)\includegraphics[width=0.95\textwidth]{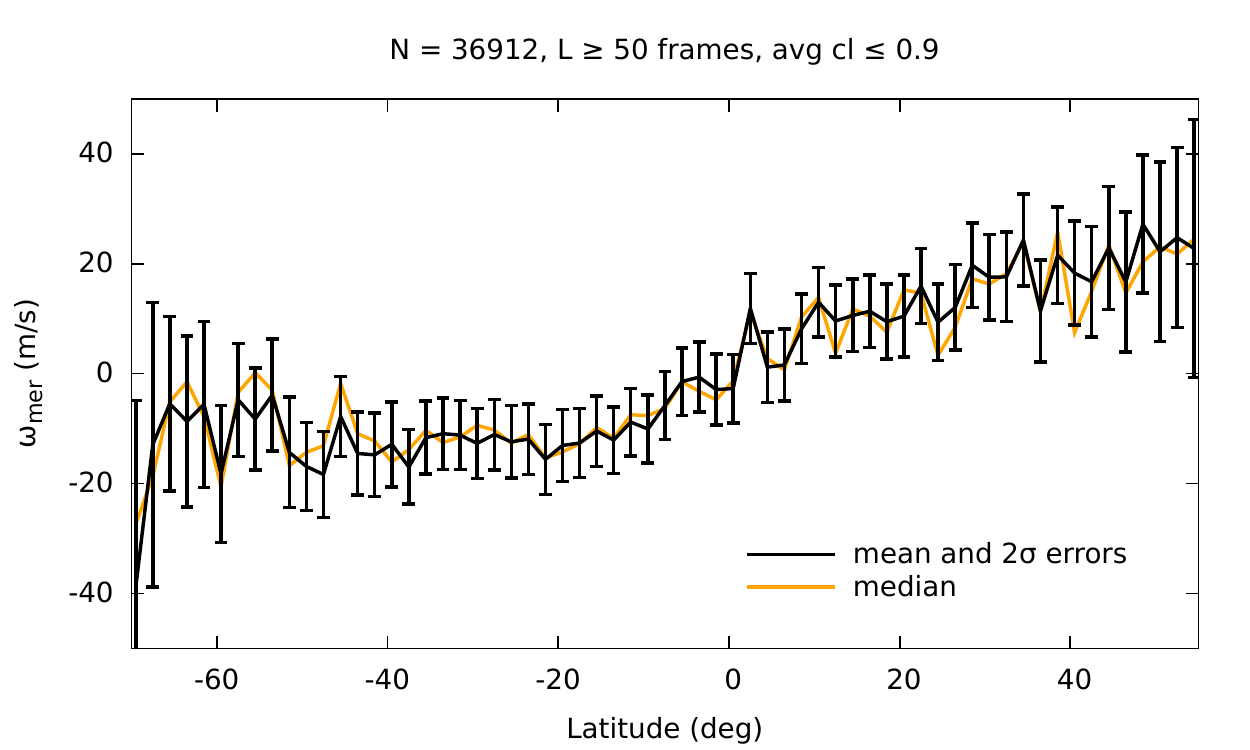}

(b)\includegraphics[width=0.95\textwidth]{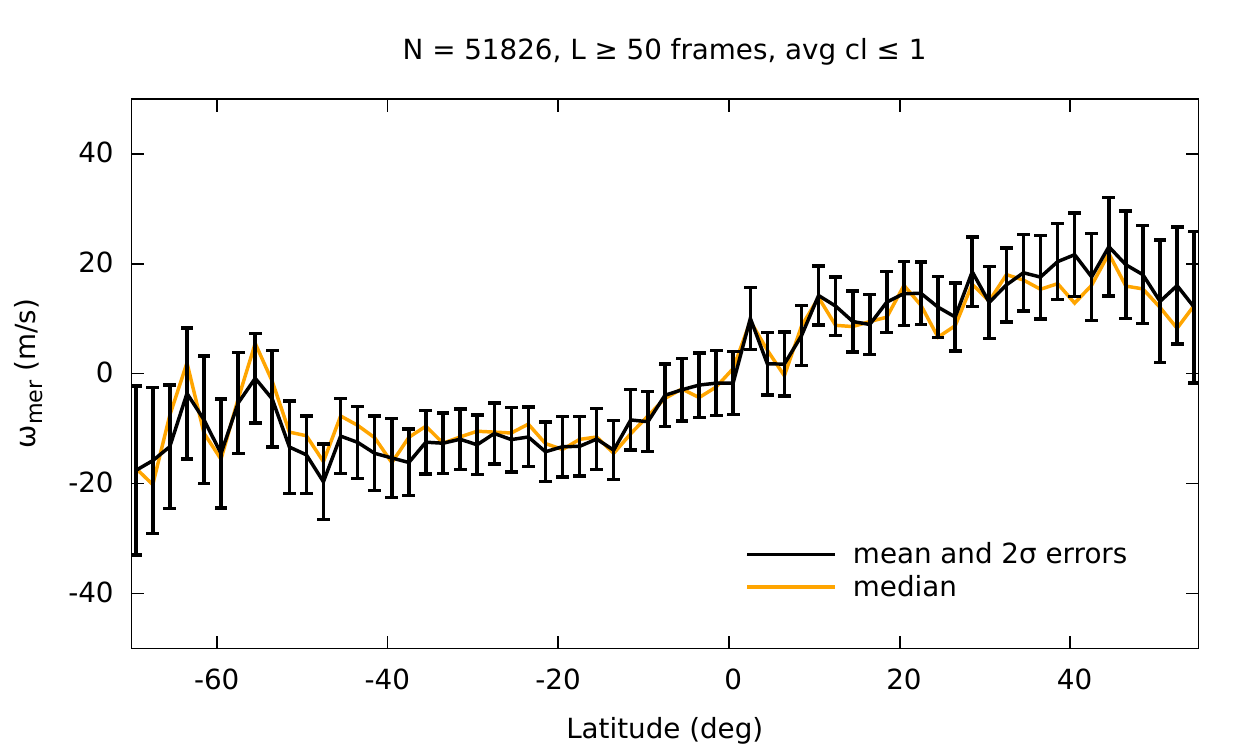}%
\end{minipage}%
\begin{minipage}[c][1\totalheight][t]{0.5\columnwidth}%
\vspace{0pt}
(c)\includegraphics[width=0.95\textwidth]{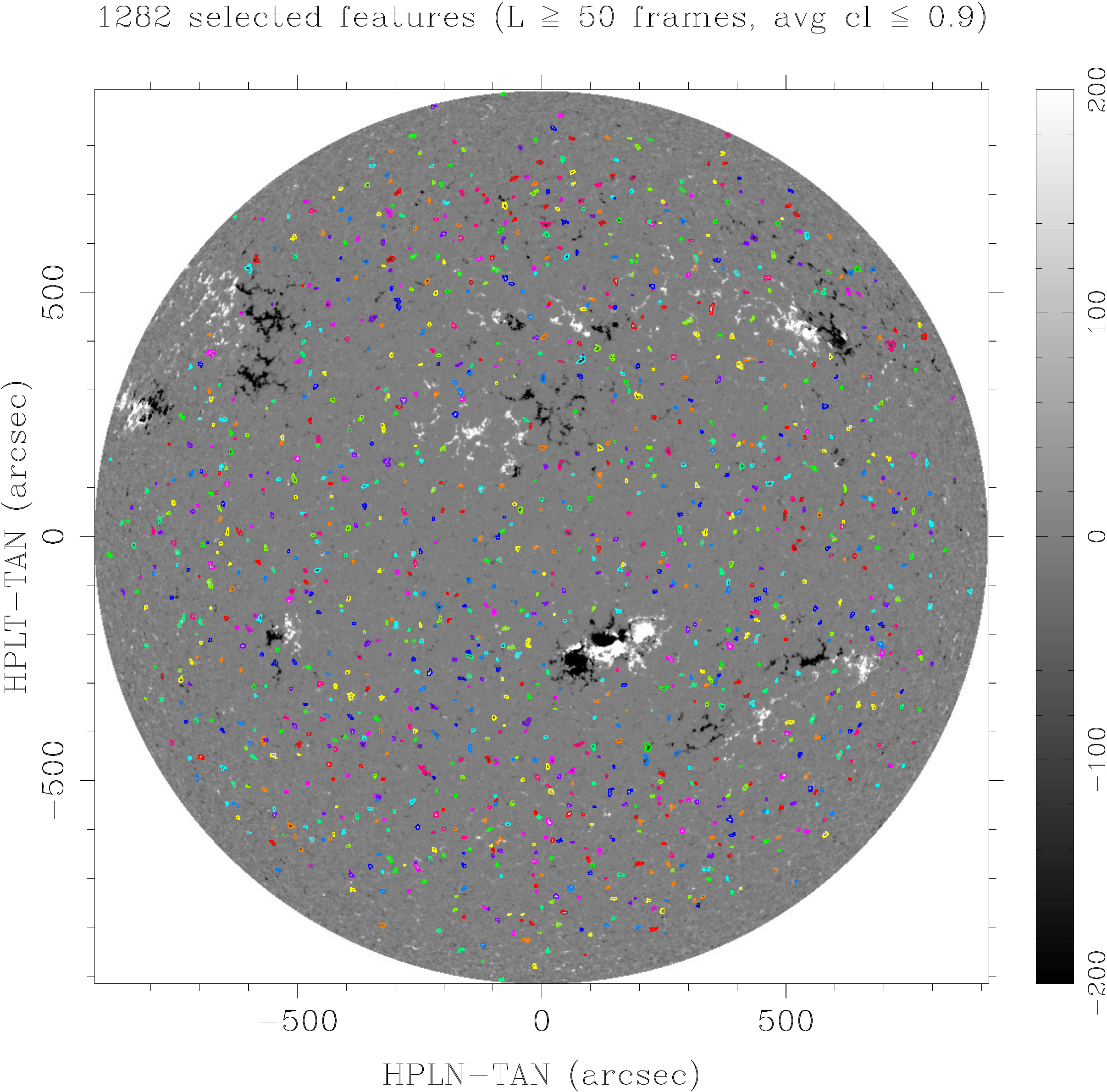}%
\end{minipage}

\caption{\label{fig:meridional-closeness-effect}Effect of the average closeness
threshold on the meridional flow profile. For a lifetime threshold
of 50~frames and average closeness of \emph{(a)} 0.9 and \emph{(b)}
1.0. Panel (b) is the same data as the right middle panel of Figure~\ref{fig:meridional-motion-for},
but with a reduced velocity axis range. \emph{(c)} Similar to Figure~\ref{fig:selected-features},
but for an average closeness of 0.9. The most obvious effect is that
the active region features are no longer selected, but a closer inspection
reveals features in the quiet sun that are excluded as well.}

\end{sidewaysfigure}

At the 50-frame lifetime threshold, I examined the effect of the average
closeness threshold. Here the effect seems to be much more pronounced
than for the rotational flow, especially for the highest northern
latitudes measured in this February dataset. In particular, the peak
at $\sim$45\degree~N shown in Figure~\ref{fig:meridional-motion-for}
disappears when the average closeness threshold is reduced from 1.0
to 0.9. Figure~\ref{fig:meridional-closeness-effect} compares the
effect of the two thresholds on the measured profile.
 I return to the effect of the choice of average closeness threshold
in \S\ref{subsec:Results-Latitudinal-Flow}.

\section{\label{sec:Results}Analytic Function Fits}

\subsection{\label{subsec:Results-Longitudinal-Flow}Longitudinal Flow}

\begin{figure}
\begin{centering}
\includegraphics[height=0.28\textheight]{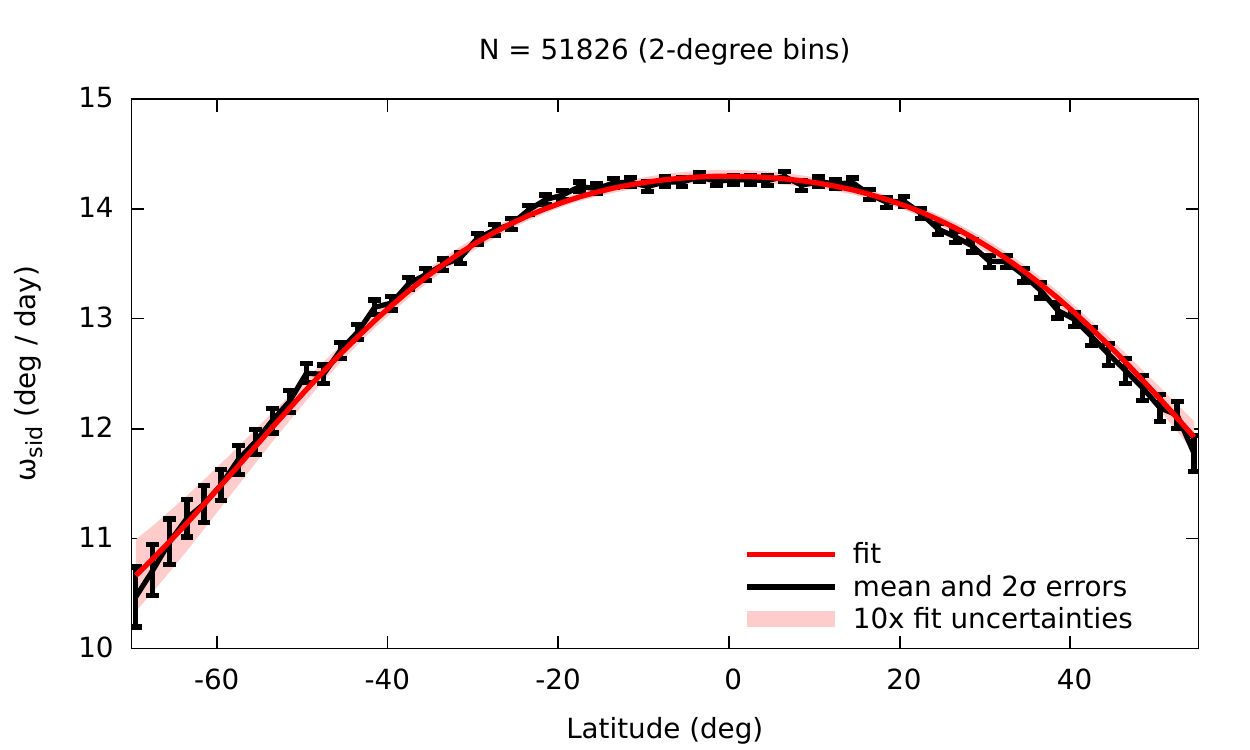}
\par\end{centering}
\begin{centering}
\includegraphics[height=0.28\textheight]{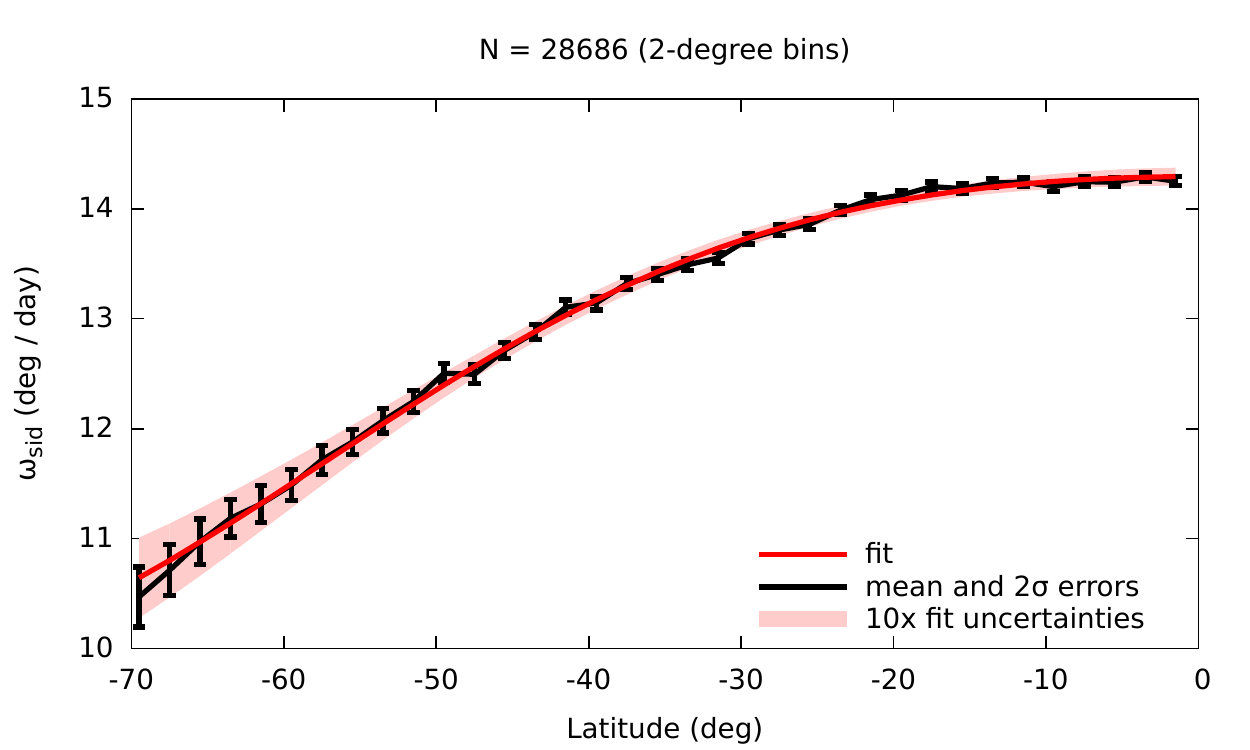}
\par\end{centering}
\begin{centering}
\includegraphics[height=0.28\textheight]{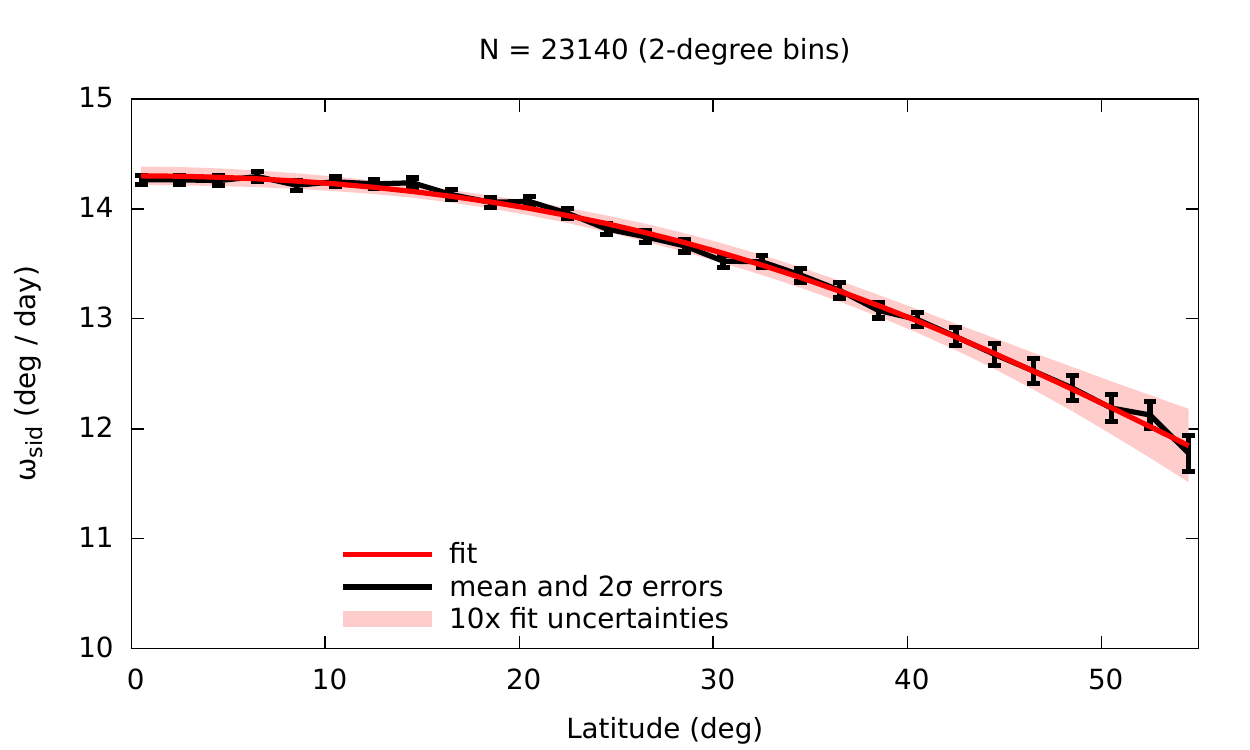}
\par\end{centering}
\caption{\label{fig:rot-fit-with-uncertainties}Differential rotation profile
for features with lifetime $\ge$50~frames and average closeness
$\le1$. Shown in each panel are (\emph{black}) the mean of the velocities
in each 2-degree latitudinal bin and error bars representing twice
the standard error of the mean, (\emph{red}) the best fit to the mean,
and (\emph{pink}) $10\times$ the uncertainties in the fit given by
$\pm10\,s(b)$ from equation~(\ref{eq:diffrot-fit-error}). Individual
panels show both hemispheres, the southern hemisphere only, and the
northern hemisphere only.}
\end{figure}

I used a Levenberg-Marquardt least squares routine to fit the sidereal
rotational velocities to the commonly-used function 
\begin{equation}
\omega_{sid}=A+B\sin^{2}b+C\sin^{4}b.\label{eq:rotation-ABC}
\end{equation}
Best-fit values of the parameters A, B, and C, as well as the covariance
matrix for these three parameters, is shown in Table~\ref{tab:Covariance-Matrix}
(the bottom triangle of the symmetric covariance matrix is not shown
for clarity). It is well-known that significant cross-talk exists
between the parameters $B$ and $C$, because $\sin^{2}b$ and $\sin^{4}b$
are not orthogonal functions. This cross-talk is evidenced by the
large (negative) value of $\sigma_{BC}^{2}$ in Table~\ref{tab:Covariance-Matrix}\textemdash an
increase in $B$ could result in a decrease in $C$ without substantially
affecting the overall quality of the fit. Figure~\ref{fig:rot-fit-with-uncertainties}
shows the mean of the observations, the fit from Table~\ref{tab:Covariance-Matrix},
and the estimate of the errors in the fit from the covariances (increased
by a factor of 10 for visibility). Errors in the fit are calculated
using a standard error propagation formula \citep[Chapter 3]{2003drea.book.....B},
which in this case is 
\begin{equation}
s_{sid}^{2}(b)=\sigma_{A}^{2}+\sigma_{B}^{2}\sin^{4}b+\sigma_{C}^{2}\sin^{8}b+2\sigma_{AB}^{2}\sin^{2}b+2\sigma_{AC}^{2}\sin^{4}b+2\sigma_{BC}^{2}\sin^{6}b,\label{eq:diffrot-fit-error}
\end{equation}
where the $\sigma_{ij}^{2}$'s are the elements of the covariance
matrix in Table~\ref{tab:Covariance-Matrix}.
\begin{table}
\caption{\label{tab:Covariance-Matrix}Best fit parameters for equation~(\ref{eq:rotation-ABC})
and parameter covariance matrices for both hemispheres, for the southern
hemisphere, and for the northern hemisphere. The covariance matrices
are symmetric; the bottom half is omitted for clarity.}

\centering{}%
\begin{tabular}{cccc}
\hline 
 & A & B & C\tabularnewline
\hline 
\hline 
Both & 14.296 & -1.847 & -2.615\tabularnewline
South & 14.292 & -1.584 & -2.938\tabularnewline
North & 14.299 & -2.124 & -2.382\tabularnewline
\hline 
\end{tabular}\qquad{}%
\begin{tabular}{cccc}
\hline 
$\sigma_{ij}^{2}(\times10^{-5})$ & A & B & C\tabularnewline
\hline 
\hline 
A & 3.38 & -23.7 & 30.4\tabularnewline
B & \textendash{} & 313 & -487\tabularnewline
C & \textendash{} & \textendash{} & 869\tabularnewline
\hline 
\end{tabular}\medskip{}
\begin{tabular}{cccc}
\hline 
$^{S}\sigma_{ij}^{2}(\times10^{-5})$ & A & B & C\tabularnewline
\hline 
\hline 
A & 6.68 & -44.0 & 53.1\tabularnewline
B & \textendash{} & 527 & -766\tabularnewline
C & \textendash{} & \textendash{} & 1270\tabularnewline
\hline 
\end{tabular}\qquad{}%
\begin{tabular}{cccc}
\hline 
$^{N}\sigma_{ij}^{2}(\times10^{-5})$  & A & B & C\tabularnewline
\hline 
\hline 
A & 7.20 & -60.2 & 91.8\tabularnewline
B & \textendash{} & 959 & -1760\tabularnewline
C & \textendash{} & \textendash{} & 3680\tabularnewline
\hline 
\end{tabular}
\end{table}
 Because of the large $B$-$C$ crosstalk, it is not appropriate to
assume that the covariances are 0, so I keep all terms of equation
(\ref{eq:diffrot-fit-error}). Using both hemispheres, the uncertainty
in the fit at 0\degree latitude is very small, only $s_{sid}(0)=\sqrt{3.38\times10^{-5}}\textrm{ deg d}^{-1}=5.8\times10^{-3}\textrm{ deg d}^{-1}$.

For comparison with prior work I also fit equation~(\ref{eq:rotation-ABC})
to the Southern and Northern hemispheres separately. \citet{Sudar2015}
found when allowing all three fit parameters to vary that the fit
parameter B was positive for both hemispheres together as well as
separately. This meant that the rotational velocity profile they found
increased slightly from the equator to $b\approx\pm18\degree$.
They did not call attention to this result\textemdash the increase
is much less than the error bars in their latitudinal bins. I find
no such effect. In all cases the fit parameter B is negative, and
the fastest rotational speed is at the equator.

\subsection{\label{subsec:Results-Latitudinal-Flow}Latitudinal Flow}

\begin{figure}
\begin{centering}
\includegraphics[height=0.28\textheight]{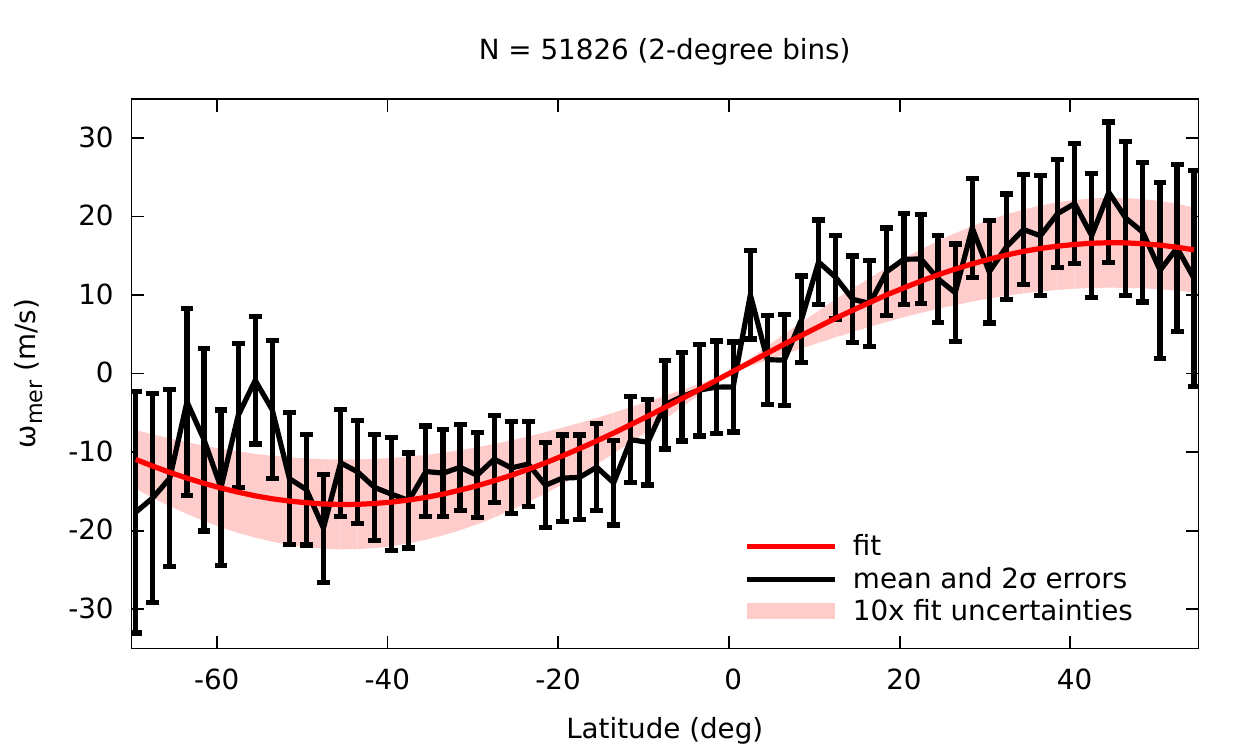}
\par\end{centering}
\begin{centering}
\includegraphics[height=0.28\textheight]{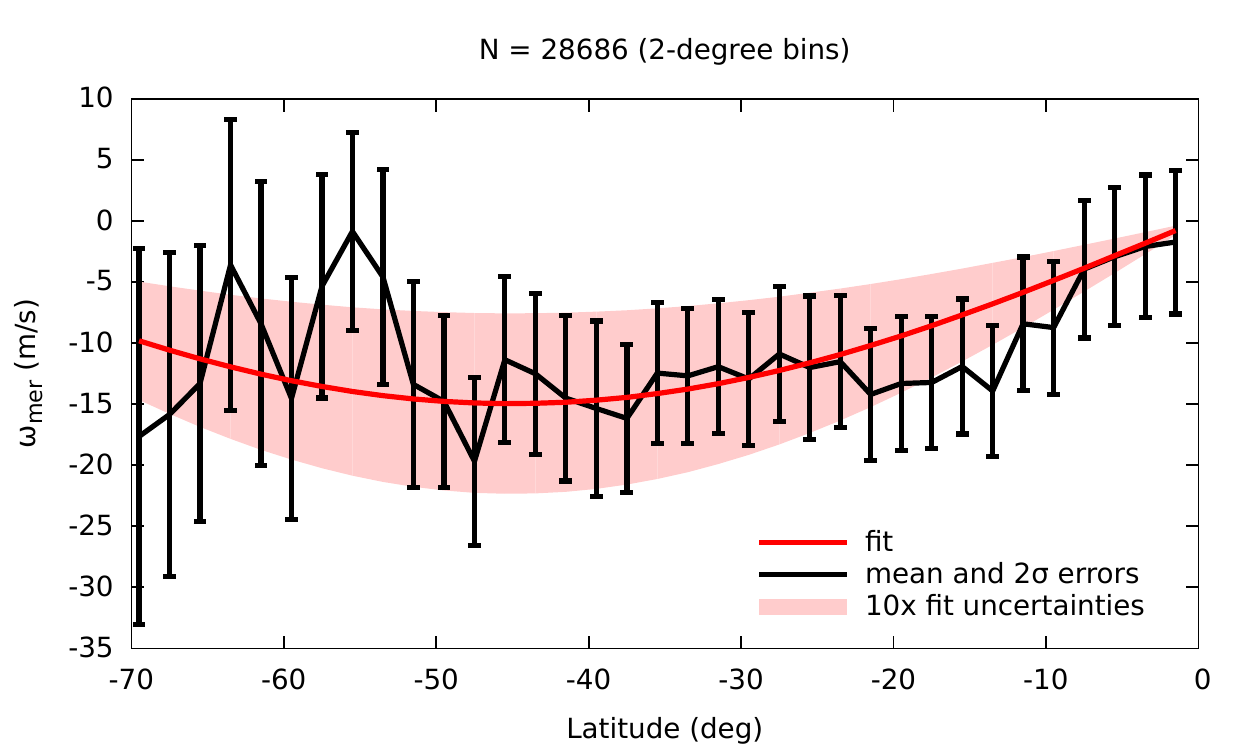}
\par\end{centering}
\begin{centering}
\includegraphics[height=0.28\textheight]{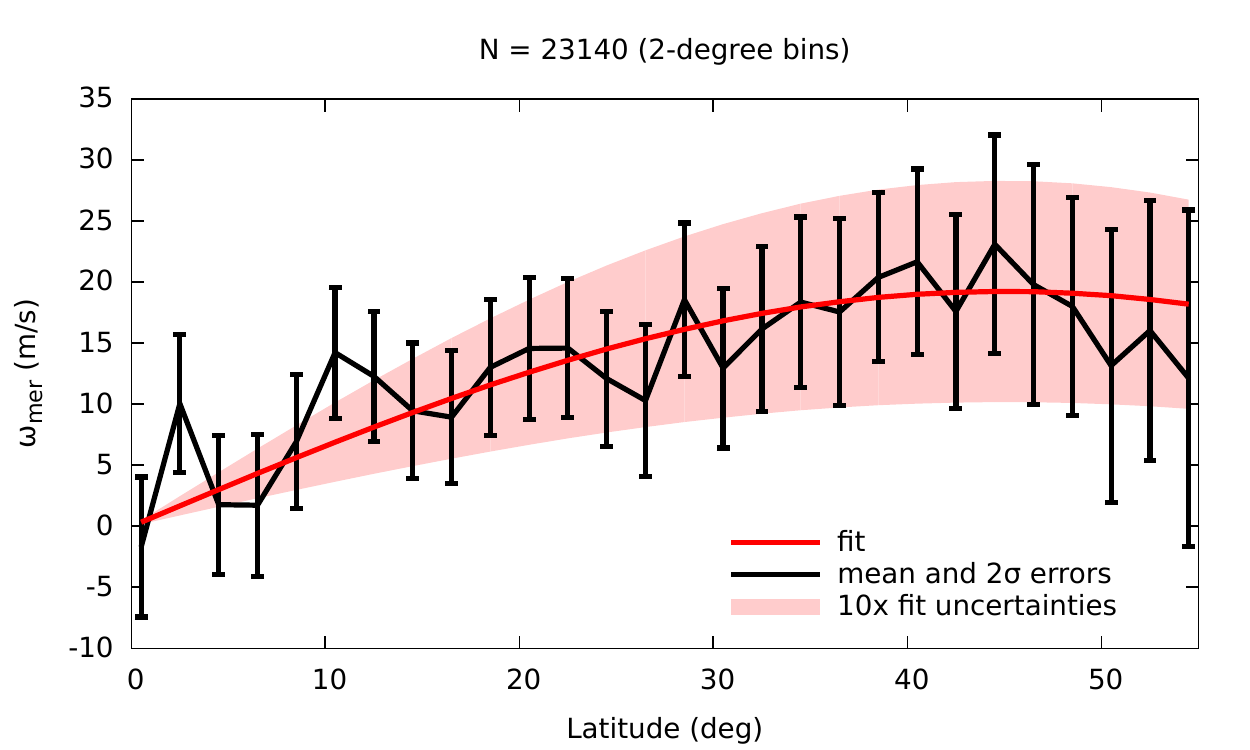}
\par\end{centering}
\caption{\label{fig:merid-fit-with-uncertainties}Meridional flow profile for
features with lifetime $\ge$50~frames and average closeness $\le1$.
Shown in each panel are (\emph{black}) the mean of the velocities
in each 2-degree latitudinal bin and error bars representing twice
the standard error of the mean, (\emph{red}) the best fit to the mean,
and (\emph{pink}) $10\times$ the uncertainties in the fit, for direct
comparison with Figure~\ref{fig:rot-fit-with-uncertainties}. Individual
panels show both hemispheres, the southern hemisphere only, and the
northern hemisphere only.}
\end{figure}

Similarly to \S\ref{subsec:Results-Longitudinal-Flow}, I use a simple
fit of the form 
\begin{equation}
\omega_{mer}=2D\sin\theta\cos\theta,\label{eq:merid-D}
\end{equation}
where $D$ is a constant with units $\text{m s}^{-1}$. This is the
dominant term by far in previous similar meridional flow measurements
\citep[e.g.][]{Hathaway2010_MeridionalFlow_SolarCycle}; given the
scatter in the measurements I do not attempt to include more terms
than this. Best fit parameters are shown in Table~\ref{tab:merid-best-fit}
for both hemispheres, the Northern hemisphere only, and the Southern
hemisphere only, and the resulting fits with uncertainties (increased
by a factor of 10 for consistency with Figure~\ref{fig:rot-fit-with-uncertainties})
are shown in Figure~\ref{fig:merid-fit-with-uncertainties}. Errors
in the fit are calculated as 
\begin{equation}
s_{mer}^{2}(b)=4\sigma_{D}^{2}\sin^{2}b\,\cos^{2}b.\label{eq:merid-fit-error}
\end{equation}
Table~\ref{tab:merid-best-fit} also shows the fit parameters for
an average closeness threshold of 0.9, corresponding to the measured
flow and selected features from Figure~\ref{fig:meridional-closeness-effect}.
\begin{table}
\caption{\label{tab:merid-best-fit}Best fit parameter for equation~(\ref{eq:merid-D})
and parameter variance for both hemispheres, for the southern hemisphere,
and for the northern hemisphere.}

\centering{}%
\begin{tabular}{ccc}
\hline 
\emph{avg cl}$\le1.0$ & D (m/s) & $\sigma_{D}^{2}$\tabularnewline
\hline 
\hline 
Both & 16.66 & 0.328\tabularnewline
South & 14.96 & 0.546\tabularnewline
North & 19.23 & 0.821\tabularnewline
\hline 
\end{tabular}%
\begin{tabular}{ccc}
\hline 
\emph{avg cl}$\le0.9$ & D (m/s) & $\sigma_{D}^{2}$\tabularnewline
\hline 
\hline 
Both & 16.69 & 0.464\tabularnewline
South & 14.82 & 0.725\tabularnewline
North & 20.00 & 1.285\tabularnewline
\hline 
\end{tabular}
\end{table}
 The small change in average closeness threshold does not significantly
change the best fit parameters, and the values for D in both cases
agree within $\pm1\sigma_{D}$.

I calculate $\chi^{2}$ for these fits as 
\begin{equation}
\chi^{2}=\sum_{i=1}^{N}\frac{(\omega_{mer,i}-\hat{\omega}_{mer,i})^{2}}{\sigma_{\mu,i}^{2}},\label{eq:chi_sq}
\end{equation}
where the index $i$ runs over the $N$ latitudinal bins, $\omega_{mer,i}$
is the mean meridional speed in that bin, $\sigma_{\mu,i}$ is the
standard error in the mean, and $\hat{\omega}_{mer,i}$ is the fit
value using equation~(\ref{eq:merid-D}) and Table~\ref{tab:merid-best-fit}.
Given only one free parameter, I estimate the number of degrees of
freedom $\nu$ as $\nu=N-1$. For the fit to the measurement in both
hemispheres with average closeness~$\le1$, $D=16.66\text{ m s}^{-1}$
and thus $\chi^{2}=91.8$ and $\chi^{2}/\nu=1.48$. For a choice of
significance level $\alpha=0.05$, the critical value of $\chi^{2}$
is $\sim80$, so at the 95\% level I must reject equation~(\ref{eq:merid-D})
as a good fit. However, I note that an outsize contribution of 13.2
to $\chi^{2}$ comes from a single latitudinal bin (-55\degree), and
were I to remove this point as an outlier I would not reject equation~(\ref{eq:merid-D})
as a good fit. (This bin is too far south to be an
artifact of active region inflow.) Instead of torturing the data
further on this point, I simply assert that in order to assess
the ability of equation~(\ref{eq:merid-D}) to describe my measurements,
more analysis at higher latitudes is necessary to both reduce the
measured uncertainties and to further constrain the trend of the mean
at these latitudes.

\section{\label{sec:Conclusions}Conclusions}

In this paper, I describe the measurement of two long-lived photospheric
flows, namely the differential rotation and meridional circulation,
in a month-long dataset. The measurements were produced directly from
the motions of both large- and small-scale \emph{individual} magnetic
features measured in high-cadence (12~min) magnetograms, rather than
by correlating thin longitudinal strips or square patches over long
time intervals. The high areal density of photospheric features and
the length of the dataset results in $7.85\times10^{6}$ tracked features.
By selecting only long-lived, relatively isolated features, I am able
to markedly reduce the effects of short-lived flows and apparent motion
due to feature interactions. Because of the large number of features
in the unfiltered dataset, this filtering still allows me to retain
excellent statistical power in the differential rotation measurements
and good power in the meridional flow measurements.

The style of my measurements is similar to the coronal bright point
rotational measurements of \citet{Sudar2015}, from which this work
has benefited greatly. Coronal bright points have the advantage of
being relatively sparse on the solar disk, and so are largely not
subject to the same feature-interaction effects as photospheric magnetic
features. I have successfully controlled for these in the case of
differential rotation, and my results are largely consistent with
past measurements. I find a rotational profile of $(14.296\pm0.006)+(-1.847\pm0.056)\sin^{2}b+(-2.615\pm0.093)\sin^{4}b$,
with units of $\textrm{ deg d}^{-1}$ and with covariances as given
in Table~\ref{tab:Covariance-Matrix}, most importantly $\sigma_{BC}^{2}=-4.87\times10^{-3}(\textrm{ deg d}^{-1})^{2}$.
I note that to fully characterize the rotational profile using equation~(\ref{eq:rotation-ABC}),
future workers must specify the full covariance matrix of the parameters
$A,B,C$, as in Table~\ref{tab:Covariance-Matrix}. Because the functions
in equation~(\ref{eq:rotation-ABC}) are not orthogonal, the variance
in the quantities B \& C over-estimates their uncertainties if the
covariance $\sigma_{BC}^{2}$ is not reported. The covariance between
the other pairs of parameters is relatively insignificant in this
work.

\begin{figure}
\begin{centering}
\includegraphics[width=1\textwidth]{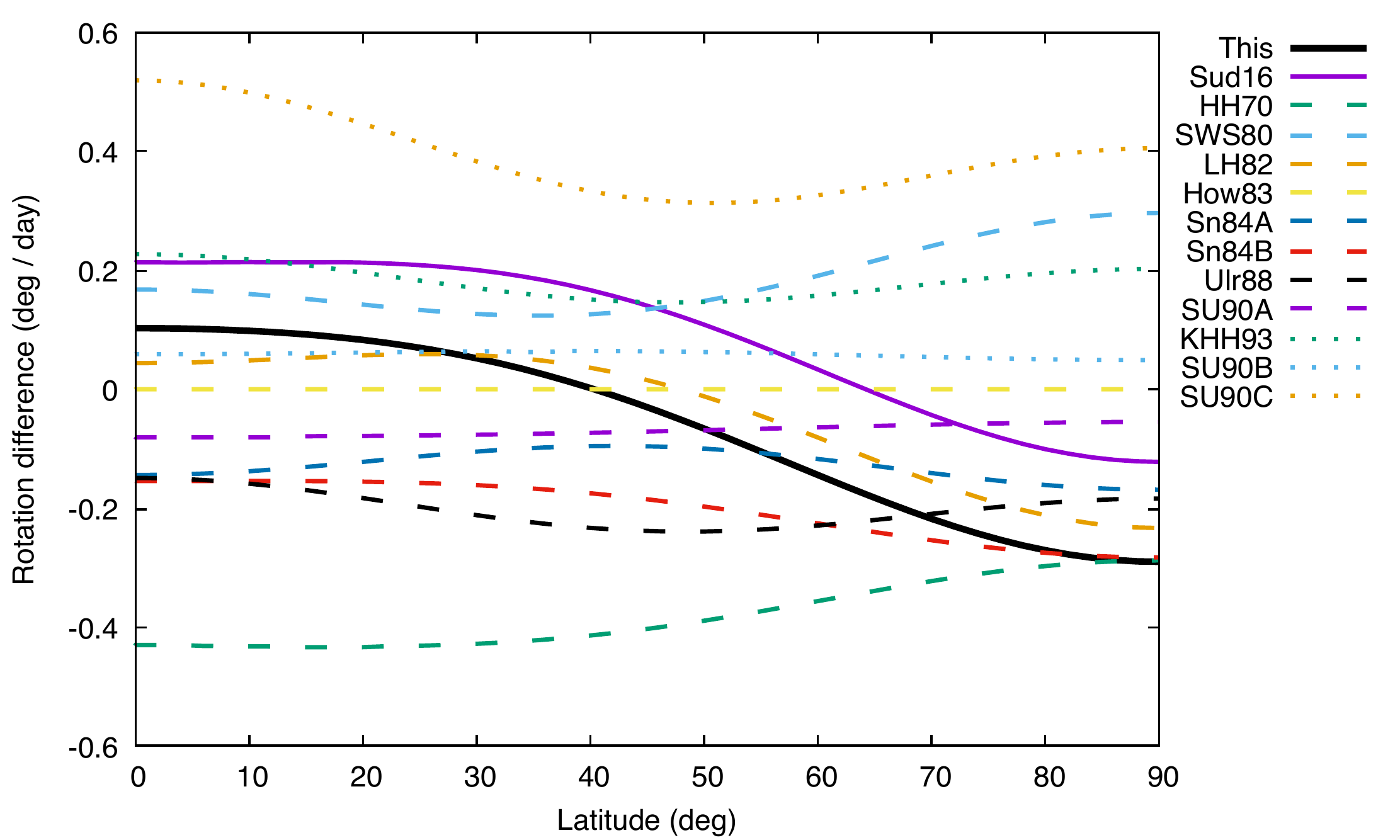}
\par\end{centering}
\caption{\label{fig:rot-comparison}Comparison of differential
rotation curves obtained from some published rotation coefficients,
in which an arbitrarily-chosen background rotation curve \citep{1983SoPh_Rotation_Mt._Wilson}
has been subtracted to emphasize the differences. Errors from uncertainties
in the coefficients have been neglected for clarity. The rotation
curve found in this work (\emph{solid black line})
is consistently slower than the coronal bright point curve of \citet{Sudar2016}
(\emph{solid purple line}). Surface
spectroscopic observations (\emph{dashed lines})
and surface tracer observations (\emph{dotted lines})
drawn from \citet{Beck2000SoPh} are also plotted. HH70: \citet{1970SoPh...12...23H};
SWS80: \citet{1980ApJ...241..811S}; LH82: \citet{1982SoPh...80..361L};
How83: \citep{1983SoPh_Rotation_Mt._Wilson}; Sn84: \citet{1984SoPh...94...13S}(A
- using all their data, B - using only their ``best'' data); Ulr88:
\citet{1988SoPh..117..291U}; SU90: \citet{1990ApJ...351..309S}(A
- spectroscopic; B - correlation of daily magnetograms; C - correlation
of supergranules in dopplergrams); KHH93: \citet{1993SoPh..145....1K}.}

\end{figure}

The solar surface rotation has been observed for
hundreds of years, and with relatively high precision for the last
50 years. \citet{Beck2000SoPh} provides a summary of results through
the end of the 20th century, and Table~2 of \citet{Sudar2015} includes
a few more recent results, including some coronal bright point measurements.
I note that their coronal bright point measurements show $14.39\le A\le14.65$,
which is consistently faster than my measured $A=14.30$.
Figure~\ref{fig:rot-comparison} compares the differential rotation
curves from this work with the larger coronal bright point dataset
of \citet{Sudar2016} as well as several spectroscopic and tracer
references drawn from \citet{Beck2000SoPh}. Again, it is obvious
that the rotation rate derived in this work is slower than that of
the coronal bright point measurement. Considering that coronal bright
points are believed to be anchored to small photospheric magnetic
features, that most of the magnetic features used in the present work
are small, and that the coronal bright points are consistently measured
to rotate faster at the equator than the magnetic features, further
work needs to be done to resolve this discrepancy. In
comparison to previous surface measurements, the equatorial rotation
rate found here is slower than two of the three tracer measurements
(and slower than the third for latitudes poleward of 30\degree). Similarly,
the equatorial rotation rate found here is faster than seven of the
eight spectroscopic rotation rates; this trend is in agreement with
previous assessments of the discrepancies between different measurement.
I note that if my measured rotation curve were to hold all the way
to the poles (where I did not in fact measure it), it would be the
slowest there among all the results in Figure~\ref{fig:rot-comparison}.

My measurement of the meridional flow is noisier than that of the
differential rotation, as expected. However, I do in all cases observe
poleward motion in both hemispheres, the amplitude of which increases
at least to latitudes of 45\degree. This is remarkable because at the
12~min cadence of my observations, and assuming a $15\text{ m s}^{-1}$
flow at 45\degree latitude, a feature would move 0.02~native HMI pixels
(0.08 pixels in my reduced-scale images) per frame. My current observations
are nearly fit well by equation~(\ref{eq:merid-D}) with a peak amplitude
of $16.7\pm0.6\text{ m s}^{-1}$, though more measurements are required
to constrain the high-latitude behavior. That the
measured flow speed increases to at least latitudes of 45\degree places
this measurement in rough agreement with other measurements of meridional
flow derived from magnetic tracking \citep[e.g.,][]{1993SoPh..145....1K,Hathaway2010_MeridionalFlow_SolarCycle}
and by extension also the coronal bright point results \citep{Sudar2016},
though I note that the peak flow amplitudes from the coronal bright
points are $\sim$50\% higher than reported here. These high latitude
peak flows are in contrast with some measurements of the meridional
flow from doppler shift, which show peak flows at lower latitudes
\citep[e.g.,][]{Ulrich2010}. \citet{Dikpati+2010} attribute the
differences in these types of measurements to the effects of surface
turbulent magnetic diffusion. I note that all measurements of meridional
circulation on the surface and in the solar interior \citep[e.g.,][]{BasuAntia2010}
show, over the course of the solar cycle, significant variations in
the latitude and amplitude of the peak flow. For example, the Mount
Wilson doppler-shift measurements show sharp low-latitude ($\sim$20\degree)
peaks in the ascending phase of the solar cycle (1987\textendash 1990,
1997\textendash 2000) and broader mid-latitude ($\sim$40\degree) peaks
during the declining phases (1992\textendash 1996, 2003\textendash 2008).
Because the descending phase of the solar cycle is longer than the
ascending phase, cycle averages tend to wash out the ascending phase
peaks and result in broader peak flows than actually exist at a given
time \citep[Figures 6 and 7]{Ulrich2010}. Thus rigorous comparisons
between measurements must take into account not only differences in
the methods, but also the averaging done for each and their relative
phase in the solar cycle.

This work shows that measuring the motions of individual
features in photospheric magnetograms can produce high precision results
in relatively short time spans. The relative ease by which this process
could be automated suggests that higher resolution non-longitudinally
averaged photospheric velocity residual measurements could be produced
to compare with coronal results, and to provide other diagnostics
of the solar dynamo.

\acknowledgements{I thank the anonymous referee for suggestions which
improved the quality of this manuscript, particularly the Introduction
and Conclusion, Ivica Skokic for detailed instructions that enabled
me to construct accurate synodic-sidereal corrections in \S\ref{subsec:The-Synodic-Sidereal-Correction},
Tim Howard for reading and providing comments on draft versions of
the manuscript, and Craig DeForest for insightful discussions regarding
the analysis. The author was partially supported by NASA Grants NNX11AP03G
and NNX14AJ67G. The data used in this paper are courtesy
of NASA/SDO and the HMI science team.}

\facility{SDO}

\bibliographystyle{aasjournal}
\bibliography{DiffRot,/Users/derek/Documents/Papers/Lamb}

\end{document}